\documentclass[a4paper,11pt]{article}
\usepackage{jheppub} 
\usepackage{orcidlink}
\usepackage{microtype}
\usepackage{physics}

\usepackage{mathtools}
\usepackage{amsmath}
\usepackage{amsfonts}
\usepackage{amssymb}
\usepackage{bbold}
\usepackage{bm}
\usepackage{bbm}
\usepackage{esint}
\usepackage{booktabs} 
\usepackage{float}

\newcommand{\bH}{{\bf H}}
\newcommand{\bB}{{\bf B}}

\newcommand{\bP}{{\bf P}}

\newcommand{\bp}{{\bf p}}
\newcommand{\be}{{\bf e}}
\newcommand{\bsigma}{{\bm\sigma}}

\newcommand{\opsi}{{\overline{\psi}}}

\long\def\exclude#1{}

\title{Single-wave solutions of the neutrino fast flavor system.
Part~II.~Weak instabilities and their resonant behavior.
}

\author[a,b,c]{Damiano F.\ G.\ Fiorillo \orcidlink{0000-0003-4927-9850}} 
\affiliation[a]{Deutsches Elektronen-Synchrotron DESY,
Platanenallee 6, 15738 Zeuthen, Germany}
\affiliation[b]{Istituto Nazionale di Fisica Nucleare (INFN), Sezione di Napoli,
Complesso Universitario di Monte Sant’Angelo, Via Cintia, 80126 Napoli, Italy}
\affiliation[c]{Gran Sasso Science Institute (GSSI), L’Aquila, Italy}

\author[d]{and Georg G.\ Raffelt
\orcidlink{0000-0002-0199-9560}}
\affiliation[d]{Max-Planck-Institut f\"ur Physik, Boltzmannstr.~8, 85748 Garching, Germany}

\abstract{Flavor instabilities in dense neutrino media trigger exponential growth of flavor waves, yet their nonlinear saturation remains poorly understood. We examine a simple proxy for this effect in the form of a single-wave solution of an axially symmetric fast flavor system. When the angular crossing is shallow and the growth rate of the instability correspondingly small, the flavor wave primarily affects resonant neutrinos that move in phase with it. The evolution of these resonant neutrinos becomes periodic, undergoing cycles of full flavor reversal. They feed power into the unstable wave, and subsequently return to their initial state, draining power back out. This new flavor pendulum captures the dynamics of weak, nearly monochromatic fast flavor instabilities. Since weakly unstable distributions always exhibit a narrow range of unstable wavenumbers, our model likely describes the earliest development of a flavor instability when it first appears. When the instability is not weak, the linear phase of a single-wave excitation does not connect to a regular nonlinear solution, unless the angle distribution consists of only two beams.
}

\begin{document}
\maketitle
\flushbottom

\section{Introduction}

In core-collapse supernovae and neutron-star mergers, the large electron-induced weak potential prevents neutrino flavor conversion \cite{Wolfenstein:1979ni}. However, the neutrino background itself can overcome this suppression: neutrino-neutrino refraction~\cite{Pantaleone:1992eq} can facilitate flavor instabilities, causing off-diagonal flavor perturbations to grow exponentially~\cite{Samuel:1993uw, Samuel:1995ri, Sawyer:2004ai, Sawyer:2008zs, Duan:2006an, Banerjee:2011fj, Chakraborty:2016lct, Izaguirre:2016gsx, Capozzi:2017gqd, Johns:2021qby}. Even without flavor violation, neutrinos can still exchange flavor through these collisionless instabilities~\cite{Duan:2010bg, Tamborra:2020cul, Volpe:2023met, Johns:2025mlm, Raffelt:2025wty}. In a recent series of papers, we have developed the general theory of flavor instabilities \cite{Fiorillo:2024bzm, Fiorillo:2024uki, Fiorillo:2024pns, Fiorillo:2025ank, Fiorillo:2025zio}. We here focus on fast flavor evolution \cite{Sawyer:2004ai, Sawyer:2008zs, Izaguirre:2016gsx, Fiorillo:2024bzm, Fiorillo:2024uki}, corresponding to massless neutrinos, the purest form of collective flavor evolution. In particular, we examine an axially symmetric setup of a single wave, where the field of flavor coherence varies with a single wave number $K$, whereas the neutrino occupation numbers remain homogeneous, leading to a closed set of equations with solutions that include traditional nonlinear flavor waves as well as new exponentially growing modes \cite{Liu:2025muc, Fiorillo:2026ybk}. In the linear phase, these solutions coincide with the usual fast flavor instabilities for a fixed wave number $K$.

In Paper I of this series \cite{Fiorillo:2026ybk}, we examined this system from the perspective of a mechanical ensemble of classical interacting spins, in analogy with the homogeneous case. One conclusion is that, in the nonlinear phase, a single wave grows into regular motion in the form of a flavor pendulum only in the special case of two angular beams. For a realistic angular distribution, however, such regularity is not protected by Gaudin invariants, in contrast to the homogeneous system. Consequently, a single-wave solution with a strong instability does not exhibit a regular nonlinear evolution.

The single-wave system can be viewed as an ensemble of interacting classical spins, in analogy with the traditional slow and fast homogeneous systems. From this perspective, the nonexistence of regular, pendulum-like solutions provides an interesting counterexample to properties that one might have taken for granted. On the other hand, this issue can also be regarded as largely academic, because strongly unstable systems are likely unphysical: they imply an inconsistent evolution toward such a state~\cite{Johns:2023jjt,Johns:2024dbe, Fiorillo:2024qbl}. For this reason, we have frequently argued that \textit{weak instabilities\/} are the most interesting ones~\cite{Fiorillo:2024qbl,Fiorillo:2024bzm,Fiorillo:2025npi}. In this case, exciting a single-wave instability is not merely a possible formulation of the problem, but could in fact be quite realistic. A weakly unstable distribution possesses only a narrow range of wavenumbers $K$ that lead to an instability~\cite{Fiorillo:2024bzm, Fiorillo:2024uki, Fiorillo:2024dik}, so one may expect that a nearly monochromatic perturbation will be the only growing mode.

A weak instability, besides favoring a single-$K$ configuration, is special in more than one way, in that only a narrow resonant range of angular modes strongly moves. The axially symmetric system depends only on the projection $v=\cos\theta$ of the neutrino velocities on the symmetry direction, and the initial configuration is represented by what we usually call~$D_v$, the difference of lepton number $(\nu_e-\bar\nu_e)-(\nu_\mu-\bar\nu_\mu)$ between the two flavors in the different $v$-beams, that are actually cones of neutrinos with equal $v$. If $D_v$ changes sign as a function of $v$, called a crossing, it is guaranteed that there is a $K$ for which the eigenfrequency $\Omega=\Omega_R+i\gamma$ has a nonvanishing imaginary part, representing an instability. If an uncrossed $D_v$ is slightly deformed to obtain a crossing, the $v$-range of what we call ``flipped neutrinos'' has $D_v$ values that are small and negative compared with typical positive ones, i.e., the population of flipped neutrinos is small compared with the unflipped one. For illustration, we are having in mind a reference distribution of the form
\begin{equation}\label{eq:reference-distribution}
    D_v=e^{-2(1-v)^2}-a,
\end{equation}
which, for $e^{-8}<a<1$, has a crossing at $v_{\rm cr}=1-\sqrt{-\log\sqrt{a}}$. For small $a$, the flipped range itself is not particularly small, but the flipped population is.

\begin{figure}
    \centering
    \includegraphics[width=\textwidth]{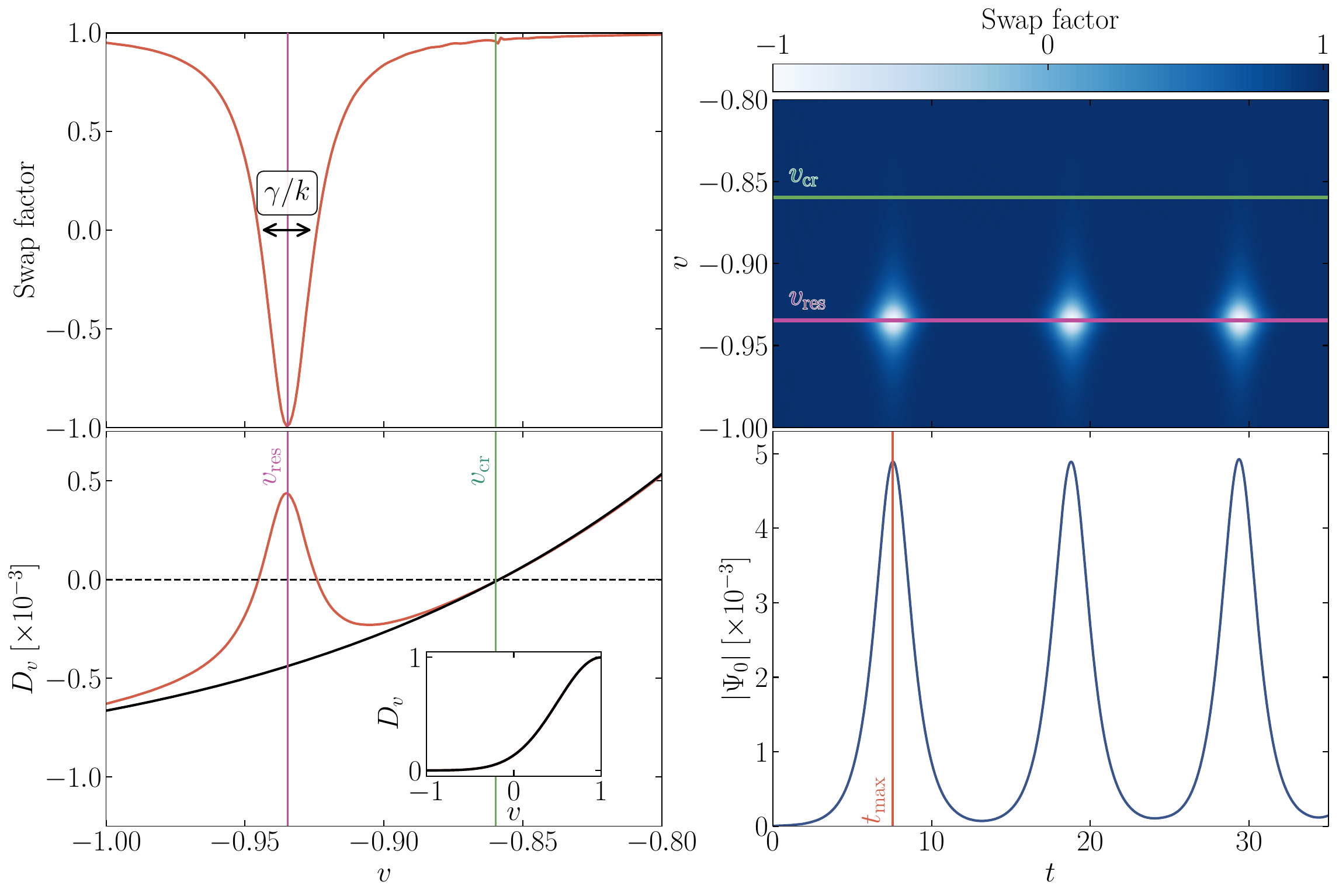}
    \vskip-6pt
    \caption{Numerical evolution of the single-wave instability for the spectrum of Eq.~\eqref{eq:reference-distribution} with $a=0.001$. \textit{Left bottom:}~Spectrum $D_v(0)$ and its maximum excursion. \textit{Left top:} Maximum swap factor, showing Lorentzian shape.
    \textit{Right bottom:} Evolution of flavor field $|\Psi_0(t)|$, showing periodic collective motion of the entire ensemble. \textit{Right top:} Contour of swap factor in the plane of $(v,t)$.}
    \label{fig:pendulum}
\end{figure}

In such a setup, the instability is dominated by the flipped population. What is more, actually it is only a narrow $v$-range around a resonant value $-1<v_{\rm res}<v_{\rm cr}$ within the flipped range that actually strongly moves. To illustrate this behavior we show an example in Fig.~\ref{fig:pendulum} from a numerical solution of the single-wave equations of motion. In the lower left panel, we show $D_v$ defined in Eq.~\eqref{eq:reference-distribution} with $a=10^{-3}$, which implies $v_{\rm cr}=-0.858$. To represent the global dynamics of the entire ensemble, we use the zeroth moment $\bP_0=\int dv\,\bP_v$, where $\bP_v$ are the usual lepton-number polarization vectors. In the fast flavor system, $\bP_0$ would be conserved, which is not true in the single-wave system. In particular, its transverse part in the form of the complex number $\Psi_0=P_0^x+iP_0^y$ measures the overall degree of flavor coherence of the entire ensemble. We show its evolution in the lower right panel of Fig.~\ref{fig:pendulum} that reveals a periodic motion, which numerically continues until the end of the simulation that we terminated after roughly 45 cycles. For our numerical treatment, we use a discretized set of 500 bins equally spaced with $\delta v=2/500$, so that we define $D_n=\sum_v D_v v^n \delta v$.

The motion of individual modes is shown in the upper right panel as a contour plot of what we call the swap factor, defined as
\begin{equation}
    \cos\theta_v(t)=\frac{D_v(t)}{D_v(0)},
\end{equation}
which is the cosine of the angle between $\bP_v(t)$ with the flavor direction. It varies between initially $+1$ and at most $-1$. We see that most modes are in this sense quiescent, whereas a narrow range in the flipped region periodically convert and return to their old position. The maximum excursion of $D_v(t)$, corresponding to the instant of maximum excursion of $|\Psi_0(t)|$, is shown in the left panels. On resonance ($v_{\rm res}=-0.935$), the polarization vector completely flips. In the upper left panel, we show the maximum swap factor, which is a near-perfect Lorentzian centered on $v_{\rm res}$ and width $\gamma/k$, where $\gamma$ is the growth rate of the instability and $k=K+D_1$ is the shifted wavenumber of the instability, with $D_1$ being the first moment of the initially unstable $D_v$. For the chosen $K=-0.709$ we find $\Omega= -0.318+i\,0.003$ so that $|\gamma/k|= 0.0105$ as indicated in Fig.~\ref{fig:pendulum}.

For the linear phase, we have long argued that such weak instabilities, those with $\gamma\ll|\Omega|$, can be understood as resonant phenomena~\cite{Fiorillo:2024bzm, Fiorillo:2024uki, Fiorillo:2025npi, Fiorillo:2025kko}. The neutrino plasma supports collective flavor waves, analogous to density waves in an electron plasma or spin waves in a magnet. These waves carry flavor, and can be emitted or absorbed by neutrinos, flipping their flavor in the process. In a $\nu_e$-dominated medium, the quanta of these waves (flavomons $\psi$) can be emitted in $\nu_\mu\to\nu_e+\psi$ and $\overline{\nu}_e\to\overline{\nu}_\mu+\psi$, and absorbed in the inverse reactions~\cite{Fiorillo:2025npi, Fiorillo:2025kko}. If those neutrinos that are capable of emitting flavomons (the resonant ones) are dominated by $\overline{\nu}_e$~or~$\nu_\mu$---flipped relative to the dominant flavor---an instability develops, driven by stimulated flavomon~emission. The resonance condition for the flavomon phase velocity to match the neutrino velocity is $\omega_R-v k=0$, where $\omega_R=\Omega_R+D_0$ is the usual shifted real part of the wave frequency and $k=K+D_1$ is the usual shifted wave vector. In the numerical example of Fig.~\ref{fig:pendulum}, the relevant quantities are $\Omega_R=-0.318$, $D_0=0.627$, $D_1=0.379$, and therefore $v_{\rm res}=-0.935$ as stated earlier. 

What is new in the dynamics illustrated by the example of Fig.~\ref{fig:pendulum} is that this behavior continues into the nonlinear phase, where the backreaction of the flavor wave modifies the original system. When the seed perturbations are monochromatic, they excite a single unstable mode, and only resonant neutrinos are strongly affected. These neutrinos remain phase-locked with the flavor field, and thus experience an effectively constant mixing term. The subsequent evolution is pendular: resonant neutrinos, which are initially flipped compared to the dominant lepton number (e.g.~$\overline{\nu}_e$ in a $\nu_e$-dominated plasma), flip their flavor completely, driven by the collective field of the plasma. In the process, they feed lepton number to the collective field of the $\nu_e$-dominated plasma, which therefore grows in amplitude, the hallmark of instability. When the flipped neutrinos have reversed entirely, and are dominated by $\overline{\nu}_\mu$, they begin the opposite process, draining the collective field of lepton number, which therefore is Landau-damped. 

The upper left panel of Fig.~\ref{fig:pendulum} affirms this picture: at its maximum excursion, the swap factor $D_v(t)/D_v(t=0)$ turns to $-1$ for the neutrinos on resonance, which flip completely, and then return back to their initial configuration. The swap factor has a characteristic Lorentzian form, well-expected in the linear regime. The key surprise is that it survives in the nonlinear phase.  This alternating emission--absorption cycle is a regular oscillation governed, to first approximation, by a flavor-pendulum equation. 

The single-wave system, therefore, has different faces. Unstable configurations, in the nonlinear regime, evolve as exact flavor pendula when there are two velocity beams, but generally this behavior does not occur for a continuous velocity distribution. On the other hand, once the instability is weak and only a narrow range of $v$-modes around a resonance strongly evolve, their behavior once more is that of a pendulum, i.e., once more the solution of the linear regime continues into the nonlinear phase as a pendulum with unexpected periodic evolution. This new flavor pendulum, to be quantified later, is not an exact solution of neutrino flavor evolution. Its dynamics is directly analogous to the quasi-regular dynamics identified by O'Neil for the evolution of near-monochromatic plasma instabilities~\cite{o1971nonlinear,o1972nonlinear}, an analogy on which we will elaborate in our discussion section. 

The regularity of the two-beam single-wave system descends from the very limited phase space and indeed, such a system must behave as a flavor pendulum \cite{TFP}. In our continuous single-wave system, the regularity again emerges from a limited range of angular modes, but here it is self-selected by a resonance condition. The resonant mechanism, intrinsic to flavor instabilities~\cite{Fiorillo:2024bzm, Fiorillo:2024uki, Fiorillo:2024pns, Fiorillo:2025ank, Fiorillo:2025npi, Fiorillo:2025zio}, naturally selects only a small range of neutrino modes to participate in the interaction. We develop an explicit model to show that the evolution is indeed pendular. Hence, this regular dynamics is likely the early saturation dynamics of weak instabilities, which involve only a narrow range of wavenumbers (near-monochromatic instability) and only a narrow group of resonant ``flipped'' neutrino modes.

To substantiate this broad-brush sketch of ideas, we begin in Sec.~\ref{sec:single_wave} by presenting the closed set of equations of motion when only a single wavenumber is excited. In Sec.~\ref{sec:flavor_pendulum}, we show that these equations admit---at least for very weak instabilities---a pendulum-like description of instability saturation. In Sec.~\ref{sec:numerical}, we confirm the emergence of this periodic behavior by numerically solving the exact equations. Here, we also show that stronger instabilities, in the single-wave regime, still lead to surprisingly regular evolution, although not pendular. Finally, in Sec.~\ref{sec:discussion} we summarize our results and discuss their implications for the evolution of an unstable neutrino plasma.

\section{Single-wave evolution}\label{sec:single_wave}

We set up the equations of motion (EoMs) for the fast-flavor system under axial symmetry, and we recall and rephrase the standard dispersion relation that follows from the linearized formulation. For a single wavenumber $K$ characterizing the spatial variation of the field of flavor coherence, we then derive a closed set of EoMs that remain valid even in the nonlinear regime. These equations imply that the occupation numbers stay homogeneous, while all components exhibit time variations that can be exponential.

\subsection{Equations of motion for the axially symmetric system}

The flavor evolution of the neutrino plasma is ruled by the usual mean-field kinetic equations for the flavor density matrix $\varrho_\bp$ for each neutrino momentum mode $\bp$ \cite{Dolgov:1980cq, Rudsky, Sigl:1993ctk, Fiorillo:2024fnl, Fiorillo:2024wej}. As in Paper~I, in a two-flavor setup, we use the parameterization
\begin{equation}
    \varrho_{\bp}
    =\frac{1}{2}\begin{pmatrix}
        f_{\bp}+D_{\bp} & \Psi_{\bp}^* \\
        \Psi_{\bp} & f_{\bp}-D_{\bp}
    \end{pmatrix},
\end{equation}
so that $\Psi_\bp$ measures the degree of flavor coherence and $D_\bp$ the difference in occupation number between $\nu_e$ and $\nu_\mu$, and analogous for antineutrinos. For fast flavor evolution (the limit of massless neutrinos), the equation for lepton number $\varrho_\bp-\overline\varrho_\bp$ is self-contained and energy no longer appears. Assuming axial symmetry, the remaining variable is velocity $v=\cos\theta$ along the symmetry axis, and we use $\varrho_v=n_\nu^{-1}\int (\varrho_\bp-\overline{\varrho}_\bp) p^2 dp /4\pi^2$. The kinetic equation becomes $ (\partial_t+v\partial_r)\varrho_v=-i[\varrho_0-v\varrho_1,\varrho_v]$, where $\varrho_n=\int dv\,v^n \varrho_v$ are the angular moments and $r$ the spatial coordinate. In component form, this is
\begin{subequations}\label{eq:exact_eom}
    \begin{eqnarray}\label{eq:exact_eom_1}
    (\partial_t+v\partial_r)D_v&=&\frac{i}{2}\bigl[(\Psi_0-v\Psi_1)\Psi^*_v-(\Psi^*_0-v\Psi^*_1)\Psi_v\bigr],
    \\ \label{eq:exact_eom_2}
    (\partial_t+v\partial_r)\Psi_v&=&\,i\,\bigl[(D_0-vD_1)\Psi_v-(\Psi_0-v\Psi_1)D_v\bigr],
\end{eqnarray}
\end{subequations}
where $D_v$, $\Psi_v$, $D_{0,1}$, and $\Psi_{0,1}$ all depend on $r$ and $t$. Writing the equation in component form brings out some general features that we have examined in Paper~I \cite{Fiorillo:2026ybk}. Of particular importance is that even in the nonlinear regime, a selfconsistent single-wave solution exists in which $\Psi_v(r,t)$ varies as $\Psi_v(t)\,e^{iK r}$, whereas $D_v(t)$ remains homogeneous, leading to a closed set of equations to be spelled out in Sect.~\ref{sec:SW} below.

\subsection{Linearized equations and dispersion relation}

The flavor coherence $\Psi_v$ is assumed to be initially small, but can grow exponentially, representing an instability that is the key ingredient of nontrivial collective flavor evolution. The instability itself can be diagnosed in the linear regime, assuming $\Psi_v$ initially so small that it does not feed back on $D_v$. We can then assume $D_v$ to be approximately constant, and seek solutions $\Psi_v(r,t)=\psi_v  e^{-i(\Omega t-Kr)}$, where the dependence on time and space is entirely encapsulated in the exponential factor. Introducing the shifted frequency $\omega=\Omega+D_0$ and wavenumber $k=K+D_1$, Eq.~\eqref{eq:exact_eom_2} yields the eigenmodes in the form
\begin{equation}\label{eq:eigenvector}
    \psi_v=D_v\frac{\psi_0-v\psi_1}{\omega-kv}.
\end{equation}
For any value of $k$, the allowed values of $\omega$ are found by imposing self-consistency for $\psi_0$ and $\psi_1$, which leads to the well-known dispersion relation for longitudinal modes (which preserve the axial symmetry of the distribution) in the form
\begin{equation}\label{eq:dispersion_longitudinal}
    (1-I_0)(1+I_2)+I_1^2=0,
\end{equation}
with
\begin{equation}
    I_n=\int \frac{D_vv^n}{\omega-kv}dv.
\end{equation}
Since $\omega$ is generically complex, the denominator is most suitably handled with a retarded prescription, i.e., for real $\omega$ we interpret the integral as the limit $\omega\to \omega+i\epsilon$. This is the Landau prescription, which we introduced for flavor waves in Ref.~\cite{Fiorillo:2023mze} and discussed extensively in Refs.~\cite{Fiorillo:2024bzm, Fiorillo:2024uki, Fiorillo:2024pns, Fiorillo:2025ank, Fiorillo:2025zio}. A solution with $\mathrm{Im}(\Omega)>0$ represents exponential growth of $\psi_v$ and hence a flavor instability.

The dispersion relation of Eq.~\eqref{eq:dispersion_longitudinal} looks quadratic in the $I_n$ integrals, but actually can be simplified. The recursion relation
\begin{equation}
    I_n=\frac{\omega I_{n-1}-D_{n-1}}{k}
\end{equation}
implies that the quadratic term cancels and Eq.~\eqref{eq:dispersion_longitudinal} is only linear in $I_n$. This simplification is connected with lepton-number conservation, which enforces the relation $\psi_1=\Omega \psi_0/K$, as shown in Refs.~\cite{Fiorillo:2024bzm,Fiorillo:2024uki}. We conclude that 
\begin{subequations}\label{eq:alternate-DR}
\begin{eqnarray}
    &&I_0=1+\frac{\Omega^2}{kK-\omega\Omega},
    \\[1ex]
    &&I_1=\frac{K\Omega}{kK-\omega\Omega},
    \\[1ex]
    &&K(1-I_0)+\Omega I_1=0,
    \\[1ex]
    &&\Omega(1+I_2)-K I_1=0
\end{eqnarray}
\end{subequations}
is a list of equivalent forms of the dispersion relation.

For weakly unstable configurations, the growth rate can be estimated explicitly by the procedure introduced in Refs.~\cite{Fiorillo:2024bzm,Fiorillo:2024uki}.
For a generically complex $\omega=\omega_R+i\gamma$, with $\gamma\ll\omega_R,k$, the integrals $I_n$ can be separated in their real and imaginary part
\begin{equation}
  I_n=P_n(\omega_R)+i\left[\partial_{\omega_R}P_n(\omega_R)\gamma+J_n(\omega_R)\right],
\end{equation}
where
\begin{equation}
    P_n(\omega_R)=\fint dv\,\frac{D_v v^n}{\omega_R-kv}
    \quad\text{and}\quad
    J_n(\omega_R)=-\frac{\pi}{|k|}D_u u^n
\end{equation}
where $u=\omega_R/k$ is the phase velocity selected by the condition of vanishing denominator and $\fint dv$ denotes the Cauchy principal value.

If we treat $\gamma$ perturbatively, we see that an eigenmode is approximately real with a frequency $\omega_R$ satisfying either of the equations
\begin{subequations}
    \begin{eqnarray}
    &&P_0(\omega_R)=1+\frac{\Omega_R^2}{kK-\omega_R\Omega_R},
    \\
    &&P_1(\omega_R)=\frac{K\Omega}{kK-\omega\Omega},
    \\
    &&K(1-P_0(\omega_R))+\Omega_R P_2(\omega_R)=0,\\
    &&\Omega_R(1+P_2(\omega_R))-K P_1(\omega_R)=0,
\end{eqnarray}
\end{subequations}
where one implies the other based on lepton number conservation. Furthermore, the first-order expansion in $\gamma$ of each of the equations yields four explicit expressions for the growth rate~\cite{Fiorillo:2024bzm,Fiorillo:2024uki,Fiorillo:2025npi}
\begin{subequations}
\begin{eqnarray}
    \gamma&=&\frac{\pi D_u (1-u U)^2}{|k|\left[\frac{\partial P_0(\omega_R)}{\partial \omega_R}(1-Uu)^2-\frac{U(2-Uu+U^2K/k)}{k}\right]},\\ 
    \gamma&=&\frac{\pi D_u u (1-U u)^2}{|k|\left[\frac{\partial P_1(\omega_R)}{\partial \omega_R}(1-Uu)^2-\frac{1+U^2K/k}{k}\right]},\\ 
    \gamma&=&\frac{\pi D_u(1-U u)}{|k|\left[\frac{\partial P_0(\omega_R)}{\partial \omega_R}-\frac{P_1(\omega_R)}{K}-U\frac{\partial P_1(\omega_R)}{\partial \omega_R}\right]},\\ \gamma&=&\frac{\pi D_u u(1-U u)}{|k|\left[\frac{\partial P_1(\omega_R)}{\partial \omega_R}-\frac{P_1(\omega_R)}{\Omega_R}-U\frac{\partial P_2(\omega_R)}{\partial \omega_R}\right]},
\end{eqnarray}    
\end{subequations}
where we set $U=\Omega_R/K$. As shown in Ref.~\cite{Fiorillo:2025npi}, these expressions for the growth rate can also be obtained by a diagrammatic approach, explicitly computing the rate of flavomon emission and absorption.

Exponential growth obtains only in the linear phase, when $\psi_v$ remains small enough to cause only negligible feedback on $D_v$. At later times, the growth of $\psi_v$ affects the occupation numbers through $D_v$, the instability turns nonlinear, and presumably saturates into a quasi-stationary state. Predicting this state remains a central challenge and has been the subject of many recent studies of how to determine it numerically and how to implement it in realistic astrophysical situations (see, e.g., Refs.~\cite{Wu:2021uvt, Nagakura:2022kic, Zaizen:2022cik, Richers:2022bkd, Goimil-Garcia:2025ozm,Fiorillo:2024qbl, Fiorillo:2025npi,Urquilla:2025idk}).

\subsection{Single wavenumber}

\label{sec:SW}

We here examine a new path to nonlinear saturation, which corresponds to exciting perturbations around the flavor-diagonal state with only a single wavenumber $K$. Nonlinear solutions with a single wavenumber were recently discovered in Ref.~\cite{Liu:2025muc} and in our Paper~I~\cite{Fiorillo:2026ybk}, we have extensively discussed why this structure leads to an exact decoupling from other wavenumbers. Here we directly assume a homogeneous $D_v(r,t)=D_v(t)$ and a single-wave variation of flavor coherence $\Psi_v(r,t)=\Psi_v(t)\,e^{iKr}$, where henceforth $\Psi_v$ stands for the time-dependent amplitude of this solution, and find the closed set of EoMs
\begin{subequations}\label{eq:eqn_motion}
    \begin{eqnarray}
    \dot D_v&=&\frac{i}{2}\bigl[(\Psi_0-v\Psi_1)\Psi^*_v-(\Psi^*_0-v\Psi^*_1)\Psi_v\bigr],
    \\
    \dot\Psi_v&=&\,i\,\bigl[(D_0-vD_1-vK)\Psi_v-(\Psi_0-v\Psi_1)D_v\bigr],
\end{eqnarray}
\end{subequations}
henceforth using the dot notation for a time derivative. Integrating the second line over $dv$ yields
\begin{equation}\label{eq:dotPsi}
    \dot\Psi_0=-iK\Psi_1,
\end{equation}
essentially encoding lepton-number conservation.

We have already shown in Paper~I that these EoMs descend from a Hamiltonian and thus conserve energy in the form
\begin{equation}\label{eq:energy}
    E=E_\Psi+E_D+E_K=\frac{1}{2}(\Psi_0\Psi_0^*-\Psi_1\Psi_1^*)+
    \frac{1}{2}(D_0^2-D_1^2)-K D_1.
\end{equation}
It falls into a piece provided by the field of flavor coherence, one from the diagonal entries of the density matrices, and one associated with the inhomogeneity introduced by the wavenumber $K$. We stress that it is quite nontrivial that the last piece would actually exist as such: as shown in Ref.~\cite{Fiorillo:2024fnl}, the refractive energy is exchanged with the neutrino kinetic energy, which cannot be expressed in terms of the flavor state of the neutrinos alone. It is a very specific property of single-wave solutions that the changing part of the kinetic energy can be expressed simply as $E_K$ in terms of the flavor state of the neutrinos.

If the velocity distribution consists of two discrete bins, which we may call two beams with velocity projections $v_1$ and $v_2$, the exact nonlinear solution of Eqs.~\eqref{eq:eqn_motion} is periodic with pendulum-like behavior \cite{Liu:2025muc, Fiorillo:2026ybk}. In contrast, we here use a continuous distribution on the interval $-1\leq v\leq 1$ and follow these equations through the nonlinear regime. Qualitatively, they describe the flavor state of individual neutrino modes ($D_v$~and~$\Psi_v$) evolving under the influence of the collective flavor field sourced by the entire plasma, encoded in $\Psi_0$ and $\Psi_1$.

The notion of exciting a single wavenumber in the neutrino plasma sounds contrived. A realistic astrophysical environment is far from controlled, and therefore, how could a single wavenumber emerge? However, it is precisely a realistic environment where an instability is not born as such, but rather develops from a stable configuration slowly driven towards instability~\cite{Johns:2023jjt, Johns:2024dbe, Fiorillo:2024qbl}. Hence, at its first appearance, an instability must always involve a narrow range of unstable wavenumbers~\cite{Yi:2019hrp, Fiorillo:2024dik}, which could be taken as monochromatic. Hence, a single-wave instability for a continuous angular distribution is quite relevant, because it presumably describes the early evolution of any instability driven by the most unstable eigenmode within a narrow range of wavenumbers. 

The likely realism of this situation is exemplified by the analogous case of electromagnetic plasma instabilities, which usually lead to a saturation of a single mode with the most unstable wavenumber, with the formation of spatially periodic structures which are experimentally observed (see Ref.~\cite{hutchinson2024kinetic} and references). With this motivation in mind, we proceed to examine a single-$K$ solution. Based on our arguments, this is a realistic choice primarily for a continuous angular distribution in the limit of a weak instability.

\section{Fast flavor instability as a resonance: Resonant flavor pendulum}
\label{sec:flavor_pendulum}

As a motivation for this study, we have already shown in Fig.~\ref{fig:pendulum} an exemplary numerical solution of the single-wave EoMs~\eqref{eq:eqn_motion} for a weakly crossed spectrum. We here proceed to explain the observed pendular motion analytically.

\subsection{Weak instability}

If the system were to contain only $\nu_e$, the flavor lepton number $D_v$ would be always positive and the absence of a crossing would prevent an instability \cite{Morinaga:2021vmc, Johns:2024bob, Fiorillo:2024bzm}. In the language of flavor waves, the stability is understood because flavomons---the excitations of the field $\Psi_0$---can only be absorbed, but not emitted. If a neutrino were to convert its flavor, it would reduce the overall electron lepton number, which however must be conserved. 

If instead $D_v$ is negative in a certain small angular region as shown in Fig.~\ref{fig:pendulum}, for instance dominated by $\overline{\nu}_e$ (``flipped'' neutrinos), these can convert through the stimulated process $\overline{\nu}_e \to \overline{\nu}_\mu + \psi$. Since the plasma is perturbed at a single wavenumber $K$, the emitted flavomons have the same $K$ and a real frequency $\Omega_R$ given by the dispersion relation. If~the flipped population is small, the mode is nearly stable, its frequency nearly real, and so only a monochromatic flavomon mode can be amplified. This resonance condition singles out those neutrinos that are kinematically able to emit these waves because they satisfy the energy-momentum constraint~\cite{Fiorillo:2025npi} $ \omega_R - k v = 0$ with $\omega_R = \Omega_R + D_0$ and $k = K + D_1$. This is the standard resonant picture of flavor conversion~\cite{Fiorillo:2024bzm, Fiorillo:2024uki, Fiorillo:2025npi}. For a single monochromatic wave, however, we can extend it into the nonlinear regime.

Most of the angular distribution consists of nonresonant neutrinos that are only weakly affected by the instability. These determine, to first approximation, both the eigenfrequency $\Omega_R$ and the mode structure of the flavomon field. Nonresonant neutrinos thus fix the global properties of the unperturbed mode, while resonant ones govern its growth rate through flavomon emission and absorption. Because the nonresonant component is mostly unaffected, the collective field can be written as $\Psi_0 = \psi_0(t)e^{-i\Omega_R t}$ and $\Psi_1 = \psi_1(t)e^{-i\Omega_R t}$ in terms of the slowly varying amplitudes $\psi_0(t)$ and $\psi_1(t)$. The EoMs then are
\begin{eqnarray}\label{eq:corotating_eom}
    \dot D_v&=&\frac{i}{2}\bigl[(\psi_0-v\psi_1)\psi_v^*-(\psi_0^*-v\psi_1^*)\psi_v\bigr],\\ \nonumber
    \dot \psi_v&=&\,i\,\bigl[(\omega_R-kv)\psi_v-(\psi_0-v\psi_1)D_v\bigr],
\end{eqnarray}
merely taken to a frame corotating with $\Omega_R$.

The mode structure remains essentially unchanged, and Eq.~\eqref{eq:dotPsi} yields $\dot\psi_0 + i K \psi_1 = 0$, so that 
\begin{equation}\label{eq:psi1psi0}
   \psi_1 = \frac{\Omega_R\psi_0}{K}+i\,\frac{\dot\psi_0}{K}.
\end{equation}
As a result of this connection, the collective dynamics is encoded entirely in the slow evolution of the amplitude $\psi_0(t)$. After replacing this expression in Eq.~\eqref{eq:corotating_eom}, a particularly instructive way to interpret the resulting equations is to introduce polarization vectors $\bP_v$, defined such that $D_v=P^z_v$ and $\psi_v=P^x_v+iP^y_v$. In terms of these variables, the equations become
\begin{equation}\label{eq:precession_eom}
    \dot\bP_v(t)=\bH_v(t)\times \bP_v(t),
\quad\text{where}\quad
  \bH_v=\begin{pmatrix}
      \psi_0(1-U v)\\
      -v\dot\psi_0/K\\
      \omega-kv
  \end{pmatrix},
\end{equation}
where we have introduced the notation $U=\Omega_R/K$. 

To write the EoMs in this form, we have made our first assumption, namely that $\psi_0$ does not precess any further and remains purely real, changing only in amplitude. In the linear regime, this holds true, since the precession at frequency $\Omega_R$ has been completely removed. The assumption that this also applies in the nonlinear regime is based on a general picture of weak instability, which implies that the original distribution is only weakly distorted in the nonlinear phase, and therefore its linear eigenmode maintains more or less the same precession frequency. (Its growth rate can, and will, change, since it depends on the properties of the resonant neutrinos which are always strongly affected by the instability~\cite{Fiorillo:2024bzm,Fiorillo:2024uki}.) So we will assume $\psi_0$ to be real, which is also confirmed in our numerical simulations.

The dynamics of neutrinos at every instant therefore is a pure precession around the field $\bH_v$. The main novelty of our treatment, so far, was to pass to a corotating frame where $\bH_v$ is slowly varying, since the frequency $\Omega_R$ was removed. Instantaneously, the precession frequency of neutrinos is 
\begin{equation}
    \omega_v=|\bH_v|=
    \sqrt{(\omega_R-kv)^2+\psi_0^2(1-Uv)^2+\dot{\psi}_0^2v^2/K^2}.
\end{equation}
The resonant neutrinos, in this treatment, emerge immediately as ``special'' because for them, $\omega_R-kv=0$, so they precess very slowly, and therefore are very sensitive to the impact of the slowly varying field $\psi_0$.

\subsection{Adiabatic approximation}

If $\bH_v$ were to evolve slowly, one could solve Eq.~\eqref{eq:precession_eom} by an adiabatic approximation. For most neutrinos, this approximation works very well: their typical precession frequency $\omega_v\sim \omega_R$ is much larger than the rate of change of $\psi_0$, which evolves over the large timescale of $\gamma^{-1}$ in the linear phase. However, for resonant neutrinos, the situation is more challenging: their precession frequency is dominated by terms of order $\omega_v\sim \psi_0$, so in the linear regime we have $\omega_v\ll \partial_t\log\psi_0\sim \gamma$. Even later, when $\psi_0$ grows to become comparable with $\gamma$, the conditions for the adiabatic approximation are formally only barely applicable, since the field changes on comparable timescale as the precession frequency.

On the other hand, for resonant neutrinos, the adiabatic approximation is actually very good for a different reason: for them, $\bH_v$ very nearly maintains the same direction while changing only in amplitude, since its $z$ component vanishes, and its $y$ component $H_{v}^y\sim \gamma\psi_0/K\ll H_{v}^x\sim \psi_0$. If $\bH_v$ always maintains the same direction, then the adiabatic approximation is actually exact, since the evolution is a precession around a fixed direction with a time-varying frequency. To use the quantum mechanical language, the Hamiltonians describing the precession of the spin at different times commute.

So the adiabatic approximation is always valid for off-resonance neutrinos, as well as for exactly on-resonance ones. Near resonance, this approach is not so precise, but may describe the qualitative behavior. We will later show that the inaccuracy for near-resonance neutrinos is not a real problem, since we may replace their exact dynamics with a matching procedure to the linear phase, in which the solution is known exactly. With this motivation, we now turn to the flavor evolution in the adiabatic approximation.

The key switch in perspective is to treat $\psi_0$ as an externally prescribed field. Of course, in reality its evolution depends on the neutrinos themselves, and we will later show how this dependence comes in. This viewpoint is essentially a nonlinear version of the flavomon picture introduced earlier~\cite{Fiorillo:2025npi}, where the flavor field is treated as a classical entity with an independent existence, whose dynamics is determined by its interaction with the individual neutrino modes. For weak instabilities, this decoupling is well-motivated, since the eigenmodes of the field are determined by the bulk of the neutrino distribution, which is only weakly affected by the instability.

In the adiabatic approximation, each polarization vector $\bP_v$ simply precesses around the instantaneous $\bH_v$ with a unit vector $\be_3=\hat\bH_v=\bH_v/\omega_v$, whereas the two orthogonal directions are
\begin{equation}
    \be_1=\frac{1}{\eta_v \omega_v}
    \begin{pmatrix}
    \psi_0(1-Uv)\delta_v\\
    -\dot{\psi}_v\delta_v/K\\
    -\eta_v^2    
    \end{pmatrix}
    \quad\text{and}\quad
    \be_2=\frac{1}{\eta_v}
    \begin{pmatrix}
        \dot\psi_0v/K\\
        \psi_0(1-U v)\\
        0
    \end{pmatrix},
\end{equation}
where $\delta_v=\omega-kv$ and $\eta_v=\sqrt{\psi_0^2(1-Uv)^2+\dot{\psi}_0^2v^2/K^2}$, so that $\omega_v^2=\eta_v^2+\delta_v^2$. It is useful, to guide the intuition, to consider the limiting cases of nonresonant neutrinos and exactly resonant ones. For the former, approximately $\be_1=(1,0,0)$, $\be_2=(0,1,0)$, and $\be_3=(0,0,1)$ holds, so they precess very rapidly around the flavor axis $\be_3$. In contrast, on resonance $\be_1=(0,0,-1)$, $\be_2=(0,1,0)$, and $\be_3=(1,0,0)$ holds, so they have a completely different dynamics, precessing fully around an axis orthogonal to the flavor one, which is the key property of the resonance.

The adiabatic solution can be formally written, in terms of the conserved length of the polarization vector, $P_v$, as 
\begin{equation}
    \bP_v=P_v\left[\cos\theta_v\be_3+\sin\theta_v(\cos \Phi_v \be_1+\sin\Phi_v \be_2)\right],
\end{equation}
where $\theta_v$ is independent of time, whereas $\Phi_v$ depends on time as $\Phi_v=\Phi^0_v+\int_0^t\omega_v(t')dt'$, describing the rapid precession. This solution is purely formal, since the values of $\theta_v$ and $\Phi^0_v$ are undetermined. To find them, we will now match the adiabatic solution with the linear solution which holds for very small $\psi_0$.

Let us select a certain time $\overline{t}$ within the linear phase, where $|\psi_v|\ll P_v$ for all neutrinos. At this instant, $\psi_0$ has a value $\opsi_0$, and the polarization vectors follow the unstable eigenmode of Eq.~\eqref{eq:eigenvector}. Expanding this explicitly, we can write the full polarization vector at $\overline{t}$ as
\begin{equation}
    \overline{\bP}_v=D^0_v\left[\frac{\opsi_0}{\theta_v^2}\left((1-Uv)\delta_v-\frac{\gamma^2 v}{K}\right),-\frac{\gamma\omega_R}{\theta_v^2}\left(1-Uv+\frac{v\delta_v}{K}\right),\sqrt{1-\frac{\overline{\eta}_v^2}{\theta_v^2}}\right],
\end{equation}
where we have introduced $\theta_v=\sqrt{\delta_v^2+\gamma^2}$, and $\overline{\eta}_v=\opsi_0\sqrt{(1-Uv)^2+\gamma^2v^2/K^2}$ is $\eta_v$ evaluated at $t=\overline{t}$. Here the $x$ and $y$ components have been obtained simply by taking the real and imaginary part of $\psi_v$ from Eq.~\eqref{eq:eigenvector}, while the $z$ component has been obtained by requesting that $|\bP_v|=D^0_v$.

We can use this polarization vector as an initial condition, and project it on $\be_1$, $\be_2$, and $\be_3$ to explicitly find the values of $\theta_v$ and $\Phi^0_v$. Notice that $\Phi^0_v$ is defined as the angle $\Phi_v$ at time $t=0$, not $t=\overline{t}$; however, since we will later take the limit $\overline{t}\to 0$, we can neglect this small difference. By explicitly taking the scalar products of $\overline{\bP}_v$ over the unit vectors at $t=\overline{t}$, we find
\begin{subequations}
\begin{eqnarray}
    \cos\theta_v&=&\frac{\delta_v}{\overline{\omega}_v}\left[\frac{\overline{\eta}_v^2}{\theta_v^2}+\sqrt{1-\frac{\overline{\eta}_v^2}{\theta_v^2}}\right]\simeq \frac{\delta_v}{\overline{\omega}_v},\\ 
    \sin\theta_v\cos\Phi^0_v&=&\frac{\overline{\eta}_v}{\overline{\omega}_v}\left[\frac{\delta_v^2}{\theta_v^2}-\sqrt{1-\frac{\overline{\eta}_v^2}{\theta_v^2}}\right]\simeq -\frac{\overline{\eta}_v\gamma^2}{\overline{\omega}_v\theta_v^2},\\
    \sin \theta_v\sin \Phi^0_v&=&-\frac{\gamma \overline{\eta}_v}{\theta_v^2}\simeq 0.
\end{eqnarray}    
\end{subequations}
Here we have introduced $\overline{\omega}_v=\sqrt{\delta_v^2+\overline{\eta}_v^2}$ as the value of $\omega_v$ at $t=\overline{t}$. Besides giving the exact expressions of the coefficients, we have also given their limiting values as $\overline{t}\to 0$ and in turn also $\opsi_0\to 0$. Within the adiabatic approximation, the values of $\theta_v$ and $\Phi^0_v$ remain constant, and therefore we can now extend our approximate adiabatic solution into the nonlinear phase, finally obtaining
\begin{equation}\label{eq:full_adiabatic_practical}
    \bP_v=D^0_v\left[\frac{\delta_v}{\overline{\omega}_v}\be_3-\frac{\overline{\eta}_v\gamma^2}{\overline{\omega}_v \theta_v^2}\left(\cos \Phi_v\be_1+\sin\Phi_v\be_2\right)\right],
\end{equation}
where $\Phi_v=\int_0^t \omega_v(t')dt'$.

\subsection{Field evolution}

So far, we have treated $\psi_0(t)$ as an externally prescribed field, although its dynamics is of course determined by its interaction with the neutrinos themselves. In principle, we may explicitly obtain $\psi_0$ from a consistency condition $\psi_0=\int dv\, \psi_v$ using the nonlinear solution $\bP_v$ from Eq.~\eqref{eq:full_adiabatic_practical}. However, it is not particularly convenient to use a consistency condition. Instead, we find it more practical to determine the evolution of $\psi_0$ from the energy that it exchanges with the neutrinos. In particular, we use the exact identity
\begin{equation}
    \partial_t |\psi_0|^2=-iK(\psi_1 \psi_0^*-\psi_1^*\psi_0).
\end{equation}
Since we have assumed $\psi_0$ to be real, an assumption that is very well verified also by our numerical solutions, we can rewrite this expression as
\begin{equation}\label{eq:energy_exchange}
    \dot\psi_0=K\int dv\,v P^y_v
\end{equation}
in terms of the $y$ component of the polarization vectors. From Eq.~\eqref{eq:full_adiabatic_practical}, we can explicitly obtain this component
\begin{equation}
    P^y_v=D^0_v\left[-\frac{\dot{\psi}_0v}{K \omega_v}\frac{\delta_v}{ \overline{\omega}_v}+\frac{v\dot{\psi}_0}{K \omega_v}\frac{\gamma^2}{\theta_v^2}\frac{\overline{\eta}_v}{\overline{\omega}_v } \frac{\delta_v}{\eta_v}\cos\Phi_v-\frac{\overline{\eta}_v}{\overline{\omega}_v}\frac{\gamma^2}{\theta_v^2}\frac{\psi_0(1-Uv)}{\eta_v}\sin \Phi_v\right].
\end{equation}
We are interested in the limit $\overline{\psi}_0\to 0$, and therefore $\overline{\eta}_v\to 0$; in this limit, the second term vanishes, since $\overline{\eta}_v \delta_v/\overline{\omega}_v\to 0$. In the first term, the combination $\delta_v/\omega_v \overline{\omega}_v$ can be approximately replaced by the principal value of $1/\delta_v$, since $\opsi_0\to 0$ and $\psi_0$ is at most as large as $\gamma\ll \omega_R, k$, as we will confirm later. Finally, in the third term, $\gamma^2/\theta_v^2$ and $\overline{\eta}_v/\overline{\omega}_v$ are both functions of $v$ strongly peaking close to $v\sim U$, and furthermore $\psi_0(1-Uv)/\eta_v\simeq 1$. Hence, Eq.~\eqref{eq:energy_exchange} can be written as
\begin{equation}
    \dot\psi_0=-\dot\psi_0\fint \frac{D^0_v v^2}{\delta_v}\,dv-\sin \Phi_u\mathcal{L},
\end{equation}
where we recall the definition of $u=\omega_R/k$, the phase velocity of the unstable wave, and we denote by $\fint$ the principal value of the integral, and 
\begin{equation}
    \mathcal{L}=\int dv D^0_v \frac{\gamma^2}{\theta_v^2}\frac{\overline{\eta}_v}{\overline{\omega}_v}.
\end{equation}
The value of the constant $\mathcal{L}$ depends sensitively on the adiabatic solution for $v$ close to resonance; unfortunately, this is exactly the range where the adiabatic approximation is not expected to hold beyond the qualitative level. However, we do not need the precise value of $\mathcal{L}$, since we will deduce it by matching the evolution of $\psi_0$ to the linear phase. In fact, by extracting $\dot\psi_0$ from this equation, we find a direct proportionality
\begin{equation}
    \dot\psi_0=-\frac{\mathcal{L}}{1+\fint \frac{D^0_vv^2}{\delta_v}dv}\sin \Phi_u.
\end{equation}
It must be accompanied by the equation for $\Phi_u$ approximately in the form
\begin{equation}
    \dot \Phi_u=\psi_0(1-U u),
\end{equation}
where we have neglected the small term proportional to $\dot{\psi}_0$ in $\omega_u$. These two equations can be combined into a single one for $\Phi_u$
\begin{equation}
    \ddot\Phi_u=-\frac{\mathcal{L}(1-Uu)}{1+\fint \frac{D^0_v v^2}{\delta_v}dv}\sin \Phi_u.
\end{equation}
For $\Phi_u\ll 1$, describing the linear phase of evolution, this equation must predict the standard exponential growth of $\Phi_u$ with growth rate $\gamma$, which allows us to obtain the precise value of $\mathcal{L}$ without any need for the adiabatic solution in the near-resonance range. Hence we find the approximate equations for the evolution in the form
\begin{equation}\label{eq:pendular_equations}
    \dot\psi_0=\frac{\gamma^2}{1-U u}\sin \Phi_u
    \quad\text{and}\quad
    \dot\Phi_u=\psi_0(1-Uu).
\end{equation}
The single equation for $\Phi_u$ finally is 
\begin{equation}\label{eq:pendular_equations-2}
    \ddot\Phi_u=\gamma^2\sin \Phi_u,
\end{equation}
which is the equation of a geometrical pendulum.

\subsection{Physical interpretation}

This derivation of the pendulum model for the weak instability has involved a large amount of complicated algebra. Therefore, let us pause and contemplate the physical meaning of the different steps. The pendulum model follows from a simple physical insight. The collective field $\Psi_0$, in the linear regime, precesses with a frequency $\Omega_R$; we assume that this remains true also in the nonlinear phase, so that the corotating $\psi_0$ evolves only in amplitude. The individual neutrino modes interact with $\psi_0$, whose amplitude changes over timescales $\gamma^{-1}$. Flavor evolution is approximately an adiabatic precession, so we have obtained an approximate expression within an externally prescribed field $\psi_0$.

To describe the dynamics of $\psi_0$ itself, we have relied on the equation describing the energy exchange between this field and the neutrinos. Specifically, we have considered the time evolution of $\psi_0$, finding that it comes from the interaction primarily with the resonant neutrinos around the velocity $v\simeq u$. The reason is simple: nonresonant neutrinos have a very rapid precession around the flavor axis, induced by the term $\omega_R-kv$ in Eq.~\eqref{eq:corotating_eom}. Hence, their polarization vectors are similar to a rapidly spinning top, whose axis is therefore rigid and difficult to move. Therefore, nonresonant neutrinos are not strongly affected by the collective field. The kinetic version of this statement is that nonresonant neutrinos move with a phase velocity different from the wave, and therefore their phase relation with the wave is constantly disrupted, impeding efficient energy exchange~\cite{Fiorillo:2024bzm}. In contrast, resonant neutrinos do not precess rapidly and therefore are much less rigid to move. 

Hence, we have shown that $\dot\psi_0$ is proportional to to the off-diagonal coherence of the resonant neutrinos $\sin \Phi_u$. In turn, the phase of the resonant neutrinos $\Phi_u$ changes because of the external field $\dot\Phi_u=\psi_0(1-Uu)$. Combining these two simple physical statements, we are immediately led to pendular dynamics; the final step is to match the pendulum parameters to the linear dynamics, where the field grows exponentially with a rate $\gamma$. 

From the pendular equations Eq.~\eqref{eq:pendular_equations}, we are now ready to extract the physical behavior. First of all,
Eq.~\eqref{eq:pendular_equations-2} admits a conserved energy
\begin{equation}\label{eq:energy_conservation_restricted}
    \mathcal{E}=\frac{\dot{\Phi}_u^2}{2}+\gamma^2\cos\Phi_u.
\end{equation}
A single swing of the pendulum can be obtained by taking $\mathcal{E}=\gamma^2$, corresponding to an initially upright pendulum. Integrating the energy conservation, we find the explicit dependence of $\Phi_u$ in a pendulum swing
\begin{equation}\label{eq:Phi_swing}
    \tan\frac{\Phi_u}{4}=e^{\gamma t}.
\end{equation}
Hence $\Phi_u$ swings from 0 at $t\to -\infty$ to $2\pi$ at $t\to+\infty$. In turn, the collective field $\psi_0$ is
\begin{equation}\label{eq:psi_swing}
    \psi_0=\frac{2\gamma}{1-Uu}\mathrm{sech}(\gamma t).
\end{equation}
Hence, the collective field at $t\to -\infty$ grows exponentially, peaking at $t=0$ with a maximum excursion $\psi_{0,\rm max}=2\gamma/(1-Uu)$ and then dropping back to 0.

The physics behind this pendular evolution is instructive as it highlights the role that resonance plays in fast flavor instabilities~\cite{Fiorillo:2024bzm, Fiorillo:2024uki}: initially the field grows exponentially, feeding on the lepton number of the resonant neutrinos, whose flavor lepton number $D^0_u \cos\Phi_u$ drops. When the resonant neutrinos have flipped completely, with $\Phi_u=\pi$, the sign of the energy exchange, which is proportional to $\sin\Phi_u$, changes, and therefore $\psi_0$ begins to drop in amplitude. The cycle is therefore inverted: the collective field is now feeding energy and lepton number into the resonant neutrinos, which therefore return to their initial lepton number. The simple solution in Eqs.~\eqref{eq:Phi_swing} and~\eqref{eq:psi_swing} describes only a single swing of the pendulum, what we have called a temporal soliton. With realistic initial conditions, the pendulum will then swing back to another cycle, with a quasi-periodic cycle of lepton number exchange between the collective field and the resonant neutrinos.

\subsection{Long-term robustness of the pendular evolution}

The collective field is driven primarily by the interaction with the resonant neutrinos. Neutrinos close to resonance are slightly detuned from the resonance condition, which leads to question whether they can accumulate a dephasing from the neutrinos exactly on resonance and therefore potentially break the pendulum solution very rapidly. In reality, it turns out that, for a sufficiently weak instability, the pendular behavior can survive for many cycles without strong dephasing, a surprising conclusion that we later confirm with our numerical cases. To understand this robustness, we now turn to consider the precession of the near-resonance neutrinos
\begin{equation}
    \dot\bP_v=\bH_v\times \bP_v,
\end{equation}
where in $\bH_v=(\psi_0(1-Uv),0,\delta_v)$ we may neglect the $y$ component, which we have stressed several times to be much smaller than the $x$ component. During each pendulum swing, $\psi_0$ behaves approximately as Eq.~\eqref{eq:psi_swing}. 

Intriguingly, this model, in which the $z$ component of the ``magnetic field'' $\bH_v$ is time-independent, while its $x$ component has a swing parametrized by the hyperbolic secant, admits an exact solution, corresponding to the so-called Rosen-Zener model~\cite{rosen1932double}. The outcome of this exact solution is the following: if a spin is subject to a Hamiltonian $H=\bB\cdot \bsigma/2$, where $\bsigma$ is the spin operator, and $\bB=(\Omega_0 \mathrm{sech}(t/T),0,\Delta)$, then if it is initially at $t\to -\infty$ in the up state along $z$, the probability of transition to the opposite spin state at $t\to +\infty$ is
\begin{equation}\label{eq:P_Rosen_Zener}
    P=\frac{\sin^2\left[\pi \Omega_0 T\right]}{\cosh^2\left[\pi \Delta T/2\right]}.
\end{equation}
We derive this expression explicitly in Appendix~\ref{sec:RZ}. Mapping these general results to our model, we obtain $\Delta=\delta_v$, $\Omega_0=2\gamma(1-Uv)/(1-Uu)$, and $T=\gamma^{-1}$. Replacing these expressions in the transition probability of the quantum spin $\bsigma$, and expanding for $v\simeq u$, we find
\begin{equation}
    P\simeq \frac{4\pi^2 U^2(v-u)^2}{(1-Uu)^2\cosh^2\left[\frac{\pi k (u-v)}{2\gamma}\right]}.
\end{equation}
This is the transition probability of the quantum spin; for the polarization vector $\bP_v$, this probability measures the deviation in the $z$ component after a full pendulum swing from its initial value before the swing. As expected, for resonant neutrinos, this is exactly 0; resonant neutrinos are enthralled in the periodic dynamics of the collective field, and therefore return exactly to their initial state. 

However, we can now also deduce that for near-resonance neutrinos, over a single swing of the pendulum, the $z$-component of the polarization vector is not exactly the original one, but suffers a small dephasing
\begin{equation}
    D_v\simeq D^0_v(1-P).
\end{equation}
Therefore, after $N$ swings of the pendulum, the dephasing reaches a level $(1-P)^N\simeq 1-NP$. The dephasing accumulates to a value of order unity when $NP\sim 1$. For neutrinos involved in the resonance, with a typical width $v-u\sim \gamma/k$, we see that $P\sim \gamma^2/k^2$ up to factors of order unity. Hence, the quasi-periodic pendulum is expected to survive for a large number of cycles, of the order of $N\sim k^2/\gamma^2$. This is a natural explanation for the long-term survival of the pendular behavior observed numerically. 

We stress, though, that this estimate is not a detailed prediction of the long-term breaking of the pendular dynamics. We have assumed, in the beginning, that the amplitude $\psi_0(t)$ remains real, i.e., that the collective field $\Psi_0$ only undergoes the corotation caused by the linear eigenfrequency $\Omega_R$, but otherwise stays in the same plane. In reality, there will be deviations of order $\gamma/\Omega_R$ from this perfect corotation, which are of the same order of magnitude as the velocity-dependent terms in $\Omega_0$. So, while our reasoning in this subsection suffices to show that the pendulum will acquire a gradual dephasing over a number of cycles of order $N\sim k^2/\gamma^2$, we cannot obtain the precise numerical coefficient. This is particularly important for the limit $K\to 0$; for a homogeneous instability, we know that the pendulum solution is exact~\cite{Padilla-Gay:2021haz, Johns:2019izj, Fiorillo:2023mze, Fiorillo:2023hlk}. This means that the changes in the corotation plane must cancel precisely the effects due to off-resonant velocity dephasing, leading to a perfect periodicity of all neutrino velocity modes. Indeed, from Eq.~(B13a) of Ref.~\cite{Fiorillo:2023hlk} we clearly see that there is indeed, in addition to the constant corotation with angular velocity $\Omega_R$, also a nonlinear precession in the azimuthal angle of the polarization vectors. After a full pendulum swing, the azimuthal angle has acquired an angular shift $\delta\varphi=2\arctan(\gamma/\Omega_R)\simeq 2\gamma/\Omega_R$, confirming that this effect exists only for weak instabilities. This angular shift precisely balances the dephasing among off-resonance velocity modes to ensure perfect pendular behavior only for $K=0$; the exact matching is protected by the conservation of the Gaudin invariants~\cite{Fiorillo:2023mze,Fiorillo:2023hlk,Fiorillo:2026ybk}.

\subsection{Summary of predictions}

In addition to a renewed understanding of the physics behind nonlinear instability relaxation, the pendulum model offers us the possibility of predicting several features of the evolution which our numerical solution in Fig.~\ref{fig:pendulum} completely verifies. In particular:
\begin{itemize}
    \item the collective field $\psi_0$ behaves as the angular momentum of a plane pendulum, since we have shown it to be proportional to the derivative of the azimuthal angle of the pendulum $\Phi_u$;
    \item there is a restricted law of energy conservation Eq.~\eqref{eq:energy_conservation_restricted}, which does not involve the full energy of the system but only the resonant mode;
    \item descending from energy conservation, the collective field has a maximum amplitude that is directly proportional to the growth rate of the initial instability 
    \begin{equation}
        \psi_{0,\rm max}=\frac{2\gamma}{1-U u}.
    \end{equation}
    This relation is far from trivial, and shows that the stronger the initial instability, the larger the produced collective field;
    \item the modes close to the resonant one evolve in the nonlinear regime following a Lorentzian structure inherited from the linear regime, with their $z$ components
    \begin{equation}
        D_v=D^0_v\left[1-\frac{1}{1+\frac{k^2(v-u)^2}{\gamma^2}}\right],
    \end{equation}
    as we derive in detail in Appendix~\ref{sec:RZ}.
\end{itemize}
All of these features do not come from numerical observations, but rather descend from our theory of the resonant flavor pendulum for weak instabilities. They are numerically verified by the weak-instability solution in Fig.~\ref{fig:pendulum}.

\section{Numerical solution for different depth of crossing}\label{sec:numerical}

We have seen that the single-wave system, for a weak instability, behaves like a flavor pendulum, meaning that the linearized solution connects to a regular nonlinear one. We here explore the behavior when the instability is not weak and the flipped population not small compared with the main beam.

\subsection{Instability pattern for fast flavor system}

\begin{figure}[ht]
\includegraphics[width=\textwidth]{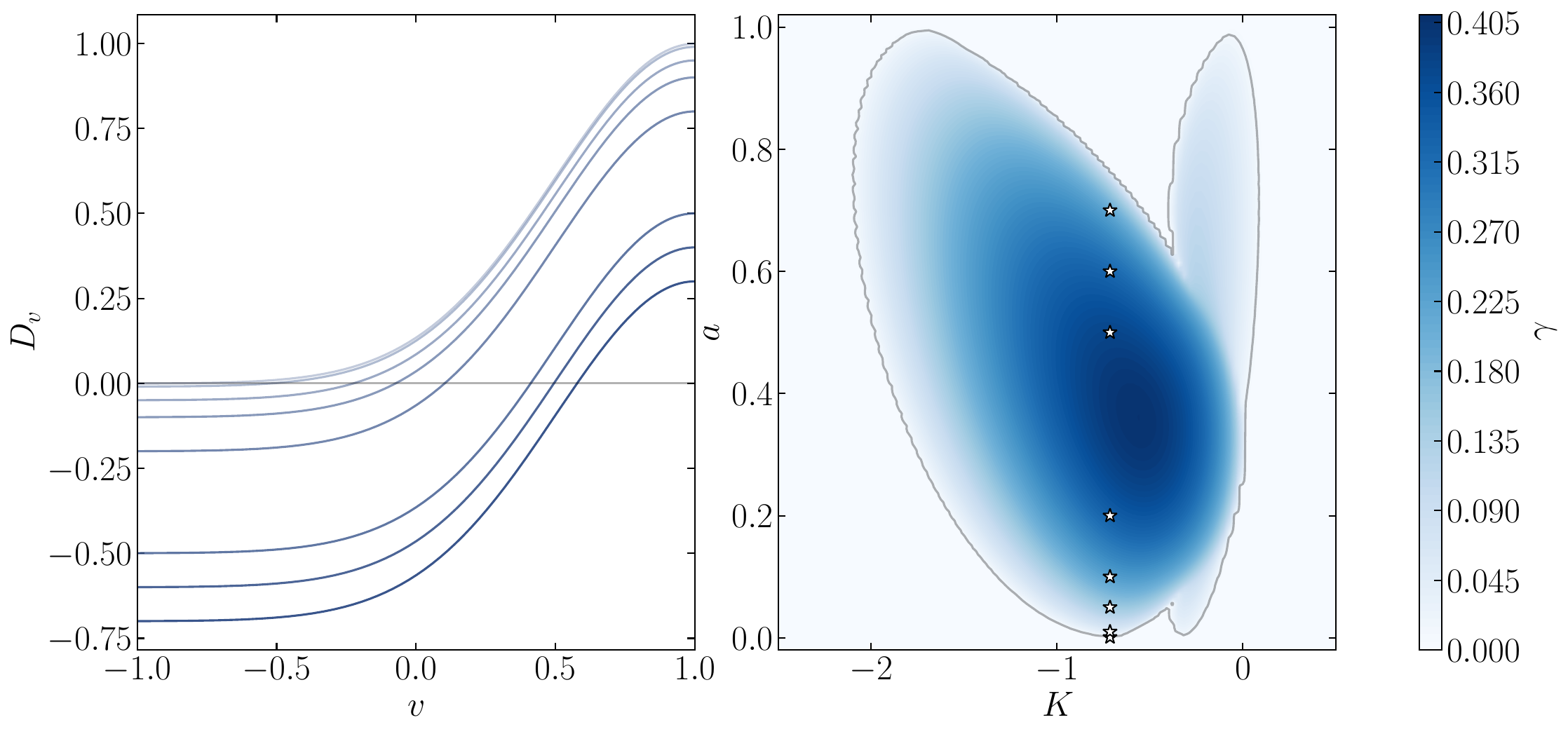}
\vskip-4pt
\caption{Instability pattern for distributions of the type Eq.~\eqref{eq:reference-distribution}. \textit{Left:} Crossed distributions for the reference values of $a$ listed in 
Eq.~\eqref{eq:avalues} and marked with asterisks in the right panel. \textit{Right:} Growth rate of unstable modes, depending on the depth of the crossing $a$ and wavenumber $K$. There is no instability outside of the thin black contour.}\label{fig:benchmark}
\end{figure}

For a more comprehensive exploration of parameter space, we use reference angular distributions of the form of Eq.~\eqref{eq:reference-distribution}. They have an angular crossing for $e^{-8}<a<1$ and in the linear regime, we may ask for the growth rate as a function of $a$ and the assumed wave number $K$, that later will be taken as a single-wave parameter. Solving the eigenvalue equation numerically for a set of 500 discrete $v$-bins, we find the contour plot shown in the right panel of Fig.~\ref{fig:benchmark}. For each $K$, there is only a single unstable eigenmode. For this discretized system, the pairs of $\{a,K\}$ within the thin black contour yield instabilities.

For very small $a$, the distribution has a very weak crossing, so that there are only two unstable wavenumbers, corresponding to the two unstable longitudinal modes (see the discussion in Refs.~\cite{Fiorillo:2024uki, Fiorillo:2024dik}). For the smallest possible $a=e^{-8}=3.35\times10^{-4}$ that yields a crossing, by modulus the larger of the two cases is $K=-0.709$, defining our reference wave number that was already used in Fig.~\ref{fig:pendulum}, where we used $a=10^{-3}$. We here use an additional list, overall considering the set of values
\begin{equation}\label{eq:avalues}
   a=\{0.001, 0.01 , 0.05 , 0.1  , 0.2  , 0.5  , 0.6  , 0.7\},
\end{equation}
denoted as asterisks in Fig.~\ref{fig:benchmark}. The angular distributions corresponding to these choices are shown in the left panel, where the top curve with $a=0.001$ is what was used for Fig.~\ref{fig:pendulum}. 

For small $a$, the forward-moving neutrinos dominate the positive difference in lepton number. As $a$ grows, the two unstable modes develop a range of unstable wavenumbers, which grow in width until the two intervals effectively merge. When $a$ becomes so large as to get comparable with unity, the crossing has flipped over completely, and the difference in lepton number is now dominated by the backward moving neutrinos. The range of unstable wavenumbers narrows to two separate regions, until the instability disappears at $a=1$.

\subsection{Single-wave nonlinear solutions}

To solve for the single-wave evolution, one can proceed in different ways. The authors of Ref.~\cite{Liu:2025muc} solved the fast flavor Eqs.~\eqref{eq:exact_eom} in coordinate space and isolated the single-wave behavior by making the box small enough that only one wavenumber was unstable. We avoid this procedure and solve directly Eqs.~\eqref{eq:eqn_motion}, which assume exactly one excited wavenumber and is a closed and self-consistent single-wave EoM. We discuss later the conditions for expecting a single-wave solution to emerge.

\begin{figure}[htbp]
    \centering
    \includegraphics[width=0.93\textwidth]{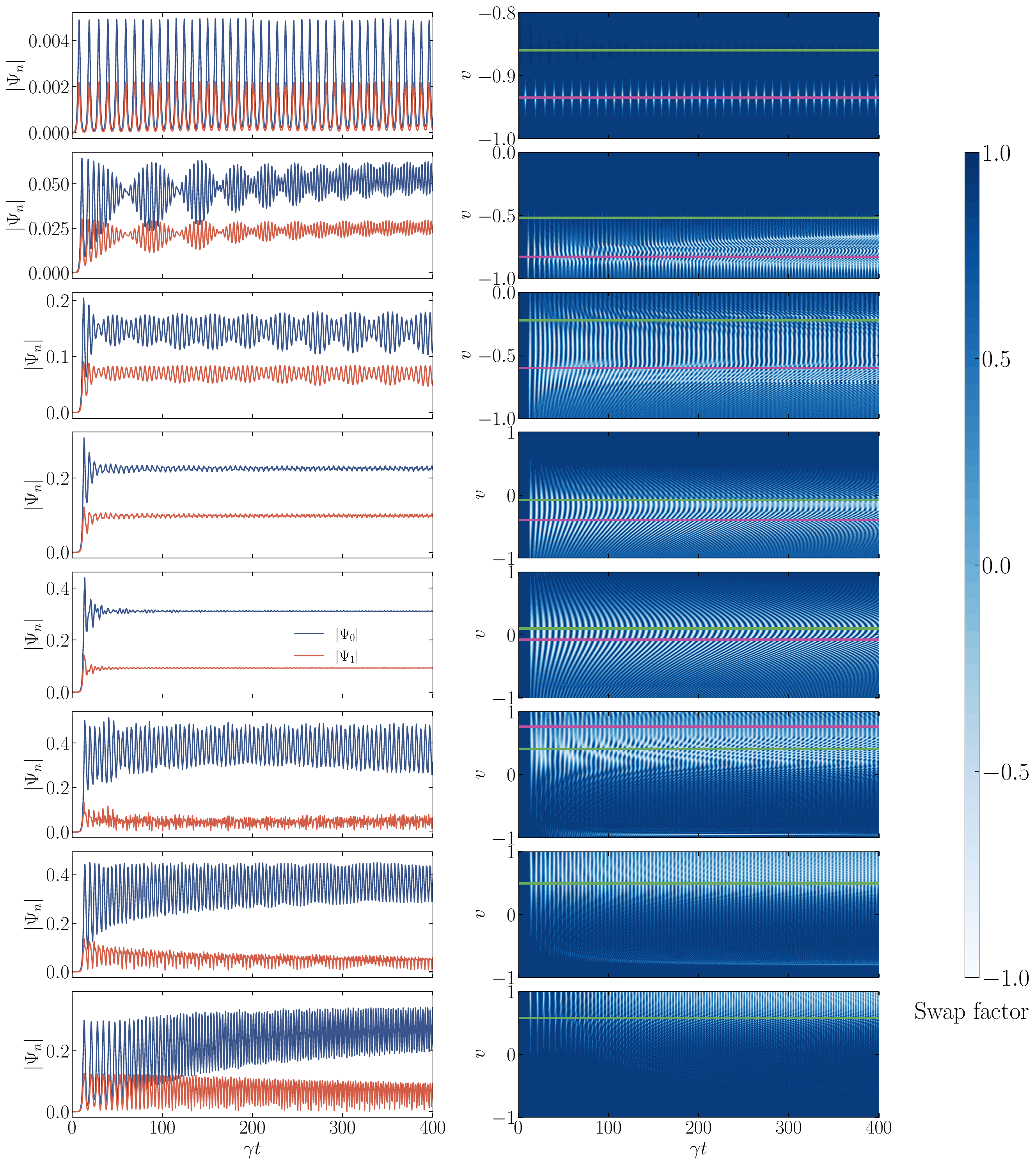}
    \caption{Time evolution for the $a$-cases listed in 
Eq.~\eqref{eq:avalues}, from top to bottom, corresponding to the angular distributions shown in Fig.~\ref{fig:benchmark}. \textit{Left:} Flavor fields $|\psi_0|$ (red) and $|\psi_1|$ (blue). \textit{Right:} Swap factor as in Fig.~\ref{fig:pendulum}. Horizontal lines denote the resonance velocity of the initial unstable mode (magenta) and the crossing velocity (green). 
    }\label{fig:time_evolution}
\end{figure}

The most interesting case, in terms of realistic emergence of single-wave solutions, is the first one, corresponding to a very weak crossing. In this case, the interval of unstable wavenumbers is very narrow, so that it is plausible to expect only a single one to be excited. In this case, a regular pendular motion emerges with characteristics shown in Fig.~\ref{fig:pendulum} that were analytically explained in Sec.~\ref{sec:flavor_pendulum}. In the top panels of Fig.~\ref{fig:time_evolution}, we show the corresponding $\psi_0(t)$ and $\psi_1(t)$ as well as the contour plot for the swap factor, but now for a much longer time. Moreover, as a time coordinate we now use $\gamma t$ and show around 45 pendular swings, a time frame over which the pendular oscillations remain regular.

In the other cases, where the crossing is deeper and the instability is stronger, the first oscillation is still reasonably described by the pendulum model, with a complete flip of neutrinos in the resonant region, although in these cases the width of the resonant region is significantly larger. For $a$ still quite small, as in the second and third row from top, the flavor fields show quasi regular behavior, suggesting beats between two regular forms of motion. In the next two rows, with much larger $a$, this behavior gives way essentially to a damping of the pendular motion and the flavor fields settling essentially to constant values. For yet larger values of $a$, we begin to approach the specular configuration, where the crossing is again shallow, but it is the forward rather than the backward modes that form a small flipped population.

The most intriguing cases are the fourth and fifth rows, corresponding to a very broad resonance and strong instability. The pendular motion is quickly damped and the system settles into a very regular solution in which $\Psi_0$ and $\Psi_1$ each perform a pure precession, with a constant modulus. This configuration is therefore strongly reminiscent of the nonlinear flavor waves discussed in Paper~I~\cite{Fiorillo:2026ybk}, where also the collective fields $\psi_0$ and $\psi_1$ perform a pure precession and so depend on time as $\Psi_n\propto e^{i(Kr-\Omega t)}$, while $D_0$ and $D_1$ constant. 

However, the nonlinear flavor waves or coprecession solutions discussed in that context were exact solutions, characterized by all the polarization vectors being in a single plane, so that all $\psi_v$ always have the same phases. The EoM for every mode is of the form $\dot\bP_v=\bH_v\times\bP_v$, and in the corotating frame, where $\psi_0$ and $\psi_1$ are static, also the $\bH_v$ are static. In the true coprecession solution, all $\bP_v$ are aligned with their $\bH_v$ and therefore also static relative to each other. On the other hand, the solutions found here still show asymptotically static $\bH_v$ in the corotating plane, but each $\bP_v$ precesses around its static $\bH_v$ with a finite-angle precession cone. However, the time-averaged $\bP_v$ are again collinear with $\bH_v$ and static relative to each other. As neighboring $\bP_v$ progressively de-phase relative to each other, the ``wiggly pattern'' seen in the right-hand panels progressively washes out. 

We have also performed analogous simulations, using a time-dependent $a(t)$, slowly increasing it from zero to different final nonzero values, similar to our setup in Ref.~\cite{Fiorillo:2024qbl}. In such simulations, one arrives at a final configuration that is close to a true coprecession solution in the sense of the $\bP_v$ being nearly co-planar. The more slowly we evolve $a(t)$, the better the coprecession solution, and the smaller the initial pendular motions. By this adiabatic deformation, one can actually transform the initially aligned polarization vectors to a pure precession mode, somewhat similar to the traditional slow-flavor spectral split phenomena~\cite{Raffelt:2007cb}, although the adiabatically changing parameter is here not the interaction strength, but the depth of the spectral crossing. Since these results are not directly relevant for the pendular evolution of weak instabilities, we do not show them here.

A final point of physical relevance is the nature of the time scales involved in the evolution. In Fig.~\ref{fig:time_evolution}, we show time in units of the individual growth rate $\gamma$ for the corresponding instability. With this choice, the pendulum oscillations, as well as the long-term non-pendular evolutions, all proceed on comparable time scales. This shows that the natural time scale of evolution, even deep in the nonlinear regime, is the growth rate of the initial instability, $\gamma$, which is far from trivial. For very weak instabilities, when the pendulum evolution applies, this descends directly from the theory laid out in Sec.~\ref{sec:flavor_pendulum}.

\section{Discussion}\label{sec:discussion}

The new approach to instability saturation studied here completely parallels the analogous one in standard electronic plasmas, that emerge when a monochromatic perturbation grows unstable. To put our results into context, it is therefore illuminating to summarize the main results of the pertinent plasma literature.

It was O'Neil and collaborators~\cite{o1971nonlinear, o1972nonlinear} who first showed that a monochromatic perturbation in an unstable plasma does not saturate quasi-linearly, but instead shows quasi-regular oscillations around a quasi-stationary state (see the very similar behavior of Fig.~1 of Ref.~\cite{o1971nonlinear}). In an electronic plasma, the origin of these oscillations is spatial trapping: an unstable wave grows and traps resonant electrons in its potential wells. Once trapping is complete and the bounce frequency of the trapped particles becomes comparable to the linear growth rate, the direction of energy exchange reverses: trapped electrons begin to transfer energy back to the wave, producing regular oscillations. The analogy with our neutrino flavor pendulum is immediate, with neutrinos exchanging energy and flavor lepton number with the flavomons.

In the neutrino context, nonlinear single-wave solutions were only recently discovered \cite{Liu:2025muc}. In a simple two-beam setup, perfectly regular evolution obtains, caused by a host of conserved quantities~\cite{Fiorillo:2026ybk}. In the generic case of a continuous angular distribution, there is no reason to expect regularity. Yet, we have uncovered a new saturation mechanism, which is generic to weak monochromatic instabilities for any angular distribution, deriving from the resonant mixing mechanism. When specialized to the two-beam case, our new mechanism does indeed reproduce the flavor pendulum introduced in Ref.~\cite{Liu:2025muc} and discussed on general grounds in Paper~I~\cite{Fiorillo:2026ybk}, but in that case, the pendulum exists for arbitrary strengths of the instability due to the limited number of degrees of freedom. Our newly discovered saturation mechanism applies only to weak instabilities, but is completely generic in that it applies to arbitrary angular distributions, and has therefore a direct application to realistic cases. 

In terms of the correspondence with conventional plasma physics, a monochromatic plasma instability does indeed lead to an electric field that grows in amplitude and then bounces back and forth, periodically exchanging energy with the population of trapped electrons moving in phase with the electrostatic wave, as seen in Fig.~1 of Ref.~\cite{o1971nonlinear}. (We somewhat disagree with the terminology of nonlinear Landau damping to describe the behavior~\cite{Liu:2025muc}, since this concept~\cite{dawson1961landau,o1965collisionless} describes the opposite case of a \textit{stable} plasma with a finite-amplitude wave superimposed.) In the same way, our newly uncovered pendulum has the collective field growing and shrinking in amplitude, periodically exchanging energy and lepton number with the resonant neutrinos, which flip completely in flavor under its action.

A crucial question is when such pendular dynamics might emerge in realistic astrophysical environments, where naively one would always expect a broadband spectrum of instabilities. On the other hand, at its first appearance, an instability is always weak, and therefore restricted to a narrow range of unstable wavenumbers, which are resonant with the narrow angular range of flipped neutrinos in the weakly-crossed distribution. Hence, understanding realistic conditions of emergence of an instability truly requires a detailed self-consistent study of astrophysical environments from the first unstable configuration. What we \textit{can} do at present is to highlight the quantitative condition that an instability should fulfill to exhibit regular dynamics.

The key requirement is that the spectrum of growing waves be so narrow that resonant neutrinos see a monochromatic wave for a timescale $\tau_{\rm nl}$ larger than a pendulum swing. The timescale for non-linear interaction descends directly from the non-linear coupling in the equations of motion, and is therefore of the order $\tau_{\rm nl}\sim |\psi_0|^{-1}$; for the pendulum model, as we have seen, the maximum amplitude is of the order $|\psi_0|\sim \gamma$, so that $\tau_{\rm nl}\sim \gamma^{-1}$. For a non-monochromatic wave, with a spread $\Delta k$, neutrinos effectively see a wavepacket with a spatial width $\Delta r\sim \Delta k^{-1}$. Since they move with the resonant velocity equal to the wave's phase velocity $v_{\rm ph}=\omega_R/k$, while the wavepacket itself moves with the group velocity $v_{\rm gr}=d\Omega_R/dK$, they remain coherent for the autocorrelation time $\tau_{\rm corr}\sim1/\Delta k|v_{\rm res}-v_{\rm gr}|$. Hence, if $\tau_{\rm nl}\ll \tau_{\rm corr}$, we should expect our regular dynamics to emerge. This reasoning completely parallels the textbook argument for bounce saturation of plasma instability~\cite{shapiro1997nonlinear}. If the instability becomes strong enough and thus non-resonant, the situation complicates, as even non-resonant neutrinos participate in the energy exchange. However, we have repeatedly argued that the evolution of the instability itself likely prevents such strong instabilities from developing. On the other hand, if the instability remains weak but $\tau_{\rm corr}\ll \tau_{\rm nl}$, the saturation regime is that of quasi-linear theory, which we introduced for the neutrino plasma in Refs.~\cite{Fiorillo:2024qbl,Fiorillo:2025npi}. A practical application of the quasi-linear approach to describe the relaxation will be studied elsewhere.

Another question is what this saturation mechanism means for slow instabilities. These are also driven by resonant flavomon emission~\cite{Fiorillo:2024pns, Fiorillo:2025ank, Fiorillo:2025zio, Fiorillo:2025kko}, and likely are the most relevant ones in SNe, since they are the first to appear~\cite{Fiorillo:2025gkw}. (The appearance of slow instabilities predating, but close to, angular crossings was also found empirically in global numerical studies~\cite{Shalgar:2022rjj, DedinNeto:2023ykt, Shalgar:2024gjt, Cornelius:2024zsb}.) On the other hand, the resonant mechanism differs as it does not single out a specific energy; rather, all antineutrinos with energy below a resonant threshold can emit flavomons~\cite{Fiorillo:2025kko}. This is a natural future direction of investigation.

\section*{Acknowledgments}

DFGF is supported by the Alexander von Humboldt Foundation (Germany). GGR acknowledges partial support by the German Research Foundation (DFG) through the Collaborative Research Centre ``Neutrinos and Dark Matter in Astro- and Particle Physics (NDM),'' Grant SFB--1258--283604770, and under Germany’s Excellence Strategy through the Cluster of Excellence ORIGINS EXC--2094--390783311.

\appendix

\section{General solution of the Rosen-Zener model}

\label{sec:RZ}

In this appendix, we discuss more generically the properties of the Rosen-Zener model \cite{rosen1932double}, and their implications for our model of the flavor pendulum associated with weak instabilities. We consider the evolution of a spin in a time-dependent magnetic field $\bB=(\Omega_0\mathrm{sech}(t/T),0,\Delta)$ governed by the Hamiltonian $H=\bB\cdot\bsigma/2$. The quantum state of the spin is $\ket{\psi}=\alpha \ket{\uparrow}+\beta\ket{\downarrow}$, so that the Schr\"odinger equation reads
\begin{eqnarray}
    i\frac{d\alpha}{dt}&=&\frac{\Delta}{2}\alpha+\frac{\Omega_0}{2}\mathrm{sech}(t/T)\beta,\\
    \nonumber i\frac{d\beta}{dt}&=&-\frac{\Delta}{2}\beta+\frac{\Omega_0}{2}\mathrm{sech}(t/T)\alpha.
\end{eqnarray}
Eliminating $\beta$, we obtain a single equation for $\alpha$
\begin{equation}
    \frac{d^2\alpha}{dt^2}+\tanh(t/T)\frac{d\alpha}{dt}+\left[\frac{\Delta^2+\Omega_0^2\mathrm{sech}^2(t/T)}{4}+i\frac{\Delta}{2}\tanh(t/T)\right]\alpha=0.
\end{equation}
We next use the substitution $\tau=[1+\tanh(t/T)]/2$, and denote by dot the derivative with respect to $\tau$ and find
\begin{equation}
    \ddot{\alpha}+\frac{1-2\tau}{2\tau(1-\tau)}\,\dot{\alpha}+\frac{\Delta^2 T^2+4\Omega_0^2 T^2 \tau(1-\tau)+2i\Delta T(2\tau-1) }{16\tau^2(1-\tau)^2}\,\alpha=0.
\end{equation}
With the substitution
\begin{equation}
    \alpha(\tau)=\left(\frac{1-\tau}{\tau}\right)^{i\Delta T/4}\gamma(\tau)
\end{equation}
we find
\begin{equation}
    \tau(1-\tau)\ddot{\gamma}+\frac{1-i\Delta T-2\tau}{2}\dot{\gamma}+\frac{\Omega_0^2 T^2}{4}\gamma=0.
\end{equation}
One solution of this equation is the hypergeometric function
\begin{equation}
    \gamma=F\left(\frac{\Omega_0 T}{2},-\frac{\Omega_0 T}{2},\frac{1-i\Delta T}{2},\tau\right).
\end{equation}
To verify that this is already the correct solution, we now test the initial conditions: for $t\to -\infty$, the wavefunction must tend to a pure spin-up state, so that $\alpha\to e^{-i\Delta t/2}$. Since for $t\to -\infty$ we have $\tau\to e^{2t/T}$, and $F[\Omega_0T/2,-\Omega_0T/2,(1-i \Delta T)/2,0]=1$, we indeed have $\alpha\to e^{-i\Delta t/2}$ as expected.

For $t\to +\infty$, we have $\tau\to 1$; in this limit, the hypergeometric function can be expressed in terms of Gamma functions, so that
\begin{equation}
    \alpha(\tau\to 1)=(1-\tau)^{\frac{i\Delta T}{4}}\frac{\left[\Gamma\left(\frac{1-i\Delta T}{2}\right)\right]^2}{\Gamma\left(\frac{1-i\Delta T+\Omega_0 T}{2}\right)\Gamma\left(\frac{1-i\Delta T-\Omega_0 T}{2}\right)}.
\end{equation}
Taking the square modulus, we obtain the probability that the spin returns to its original state $|\alpha|^2=1-P$; using known identities from the theory of Gamma functions, this can be reduced to Eq.~\eqref{eq:P_Rosen_Zener} in the main text.

It is more interesting to consider the generic time-dependent solution for the near-resonance neutrinos. Using the mapping for the parameters deduced in the main text $\Delta T=\delta_v/\gamma$ and $\Omega_0 T=2(1-U v)/(1-U u)$, for neutrinos close to resonance, we can approximate $\Omega_0 T\simeq 2$, since the off-resonance correction is small. Instead, in $\Delta T$, we have to keep of course the first-order correction $\Delta T=\delta_v/\gamma$. The solution for $\gamma$ then reads
\begin{equation}
    \gamma=F(1,-1,\frac{1-i\delta_v/\gamma}{2},\tau)=1-\frac{2\tau}{1-i\delta_v/\gamma}.
\end{equation}
Hence, the $z$ component of the spin $S^z=(2|\alpha|^2-1)/2=\cos\theta/2$, where $\theta$ is the angle formed by the spin with the $z$ axis, turns out to be
\begin{equation}
    \cos \theta=1-\frac{2\,\mathrm{sech}^2\gamma t}{1+\delta_v^2/\gamma^2}.
\end{equation}
Hence, we see that at the maximum excursion of the pendulum the effect on the neutrinos exhibits the characteristic resonant Lorentzian, with the same denominator as in the linear regime $\delta_v^2+\gamma^2$, even though we are deep in the nonlinear regime. This surprising prediction is validated by our numerical solution in the main text.

\bibliographystyle{JHEP}
\bibliography{References}

@misc{TFP,
    author = "Fiorillo, Damiano F. G. and Raffelt, Georg G.",
    title = "{The Ubiquitous Flavor Pendulum}",
    note = "{Work in progress (2026)}"
}

@article{Urquilla:2025idk,
    author = "Urquilla, Erick and Johns, Lucas",
    title = "{Testing common approximations of neutrino fast flavor conversion}",
    eprint = "2510.23917",
    archivePrefix = "arXiv",
    primaryClass = "astro-ph.HE",
    month = "10",
    year = "2025"
}

@article{Fiorillo:2026ybk,
    author = "Fiorillo, Damiano F. G. and Raffelt, Georg G.",
    title = "{Single-wave solutions of the neutrino fast flavor system. Part I. Mechanical properties}",
    eprint = "2601.15372",
    archivePrefix = "arXiv",
    primaryClass = "hep-ph",
    month = "1",
    year = "2026"
}

@article{rosen1932double,
  title="{Double Stern-Gerlach experiment and related collision phenomena}",
  author={Rosen, Nathan and Zener, Clarence},
  journal={Phys. Rev.},
  volume={40},
  number={4},
  pages={502},
  year={1932},
  doi="10.1103/PhysRev.40.502"
}

@article{Volpe:2023met,
    author = "Volpe, M. Cristina",
    title = "{Neutrinos from dense environments: Flavor mechanisms, theoretical approaches, observations, and new directions}",
    eprint = "2301.11814",
    archivePrefix = "arXiv",
    primaryClass = "hep-ph",
    doi = "10.1103/RevModPhys.96.025004",
    journal = "Rev. Mod. Phys.",
    volume = "96",
    number = "2",
    pages = "025004",
    year = "2024"
}

@article{Richers:2022bkd,
    author = "Richers, Sherwood and Duan, Huaiyu and Wu, Meng-Ru and Bhattacharyya, Soumya and Zaizen, Masamichi and George, Manu and Lin, Chun-Yu and Xiong, Zewei",
    title = "{Code comparison for fast flavor instability simulations}",
    eprint = "2205.06282",
    archivePrefix = "arXiv",
    primaryClass = "astro-ph.HE",
    reportNumber = "N3AS-22-009",
    doi = "10.1103/PhysRevD.106.043011",
    journal = "Phys. Rev. D",
    volume = "106",
    number = "4",
    pages = "043011",
    year = "2022"
}

@article{Wu:2021uvt,
    author = "Wu, Meng-Ru and George, Manu and Lin, Chun-Yu and Xiong, Zewei",
    title = "{Collective fast neutrino flavor conversions in a 1D box: Initial conditions and long-term evolution}",
    eprint = "2108.09886",
    archivePrefix = "arXiv",
    primaryClass = "hep-ph",
    doi = "10.1103/PhysRevD.104.103003",
    journal = "Phys. Rev. D",
    volume = "104",
    number = "10",
    pages = "103003",
    year = "2021"
}

@article{Goimil-Garcia:2025ozm,
    author = "Goimil-Garc{\'\i}a, Manuel and Tamborra, Irene",
    title = "{Steady state of fast-oscillating neutrinos in an inhomogeneous medium}",
    eprint = "2509.22805",
    archivePrefix = "arXiv",
    primaryClass = "astro-ph.HE",
    doi = "10.1103/gdg9-rzns",
    journal = "Phys. Rev. D",
    volume = "112",
    number = "10",
    pages = "103011",
    year = "2025"
}

@article{Raffelt:2025wty,
    author = "Raffelt, Georg G. and Janka, Hans-Thomas and Fiorillo, Damiano F. G.",
    title = "{Neutrinos from core-collapse supernovae}",
    eprint = "2509.16306",
    archivePrefix = "arXiv",
    primaryClass = "astro-ph.HE",
    month = "9",
    year = "2025"
}

@article{hutchinson2024kinetic,
  title={Kinetic solitary electrostatic structures in collisionless plasma: Phase-space holes},
  author={Hutchinson, IH},
  journal={Rev. Mod. Phys.},
  volume={96},
  number={4},
  pages={045007},
  year={2024},
  doi={10.1103/RevModPhys.96.045007}
}

@article{Fiorillo:2025kko,
    author = "Fiorillo, Damiano F. G. and Raffelt, Georg G.",
    title = "{Lepton number crossings are insufficient for flavor instabilities}",
    eprint = "2507.22987",
    archivePrefix = "arXiv",
    primaryClass = "hep-ph",
    month = "7",
    year = "2025"
}

@article{o1965collisionless,
  title="{Collisionless Damping of Nonlinear Plasma Oscillations}",
  author={O'Neil, Thomas},
  journal={Phys. Fluids},
  volume={8},
  number={12},
  pages={2255--2262},
  year={1965},
  publisher={AIP Publishing},
 doi = {10.1063/1.1761193},
}

@article{dawson1961landau,
  title="{On Landau Damping}",
  author={Dawson, John},
  journal={Phys. Fluids},
  volume={4},
  number={7},
  pages={869--874},
  year={1961},
  publisher={AIP Publishing},
  doi="10.1063/1.1706419"
}

@article{shapiro1997nonlinear,
       author = {{Shapiro}, V.~D. and {Sagdeev}, R.~Z.},
        title = "{Nonlinear wave-particle interaction and conditions for the applicability of quasilinear theory}",
      journal = {Phys. Rep.},
         year = 1997,
        month = apr,
       volume = {283},
       number = {1},
        pages = {49-71},
          doi = {10.1016/S0370-1573(96)00053-1}
}

@article{o1972nonlinear,
  title="{Nonlinear Interaction of a Small Cold Beam and a Plasma. Part II}",
  author={O'Neil, T. M. and Winfrey, J. H.},
  journal={Phys. Fluids},
  volume={15},
  number={8},
  pages={1514--1522},
  year={1972},
  publisher={AIP Publishing},
  doi="10.1063/1.1694117"
}

@article{o1971nonlinear,
  title="{Nonlinear Interaction of a Small Cold Beam and a Plasma}",
  author={O'Neil, T. M. and Winfrey, J. H. and Malmberg, J. H.},
  journal={Phys. Fluids},
  volume={14},
  number={6},
  pages={1204--1212},
  year={1971},
  publisher={AIP Publishing},
  doi="10.1063/1.1693587"
}

@article{Fiorillo:2025gkw,
    author = "Fiorillo, Damiano F. G. and Janka, Hans-Thomas and Raffelt, Georg G.",
    title = "{Neutrino-Mass-Driven Instabilities as the Earliest Flavor Conversion in Supernovae}",
    eprint = "2507.22985",
    archivePrefix = "arXiv",
    primaryClass = "hep-ph",
    doi = "10.1103/jbmx-rbzt",
    journal = "Phys. Rev. Lett.",
    volume = "135",
    number = "23",
    pages = "231003",
    year = "2025"
}

@article{Liu:2025muc,
    author = "Liu, Jiabao and Johns, Lucas and Nagakura, Hiroki and Zaizen, Masamichi and Yamada, Shoichi",
    title = "{Dynamical equilibria of fast neutrino flavor conversion}",
    eprint = "2509.26418",
    archivePrefix = "arXiv",
    primaryClass = "astro-ph.HE",
    month = "9",
    year = "2025"
}

@article{Fiorillo:2025zio,
    author = "Fiorillo, Damiano F. G. and Raffelt, Georg G.",
    title = "{Dispersion relation of the neutrino plasma: Unifying fast, slow, and collisional instabilities}",
    eprint = "2505.20389",
    archivePrefix = "arXiv",
    primaryClass = "hep-ph",
    month = "5",
    year = "2025"
}

@article{Fiorillo:2024pns,
    author = "Fiorillo, Damiano F. G. and Raffelt, Georg G.",
    title = "{Theory of neutrino slow flavor evolution. Part I. Homogeneous medium}",
    eprint = "2412.02747",
    archivePrefix = "arXiv",
    primaryClass = "hep-ph",
    doi = "10.1007/JHEP04(2025)146",
    journal = "JHEP",
    volume = "04",
    pages = "146",
    year = "2025"
}

@article{Johns:2025mlm,
    author = "Johns, Lucas and Richers, Sherwood and Wu, Meng-Ru",
    title = "{Neutrino Oscillations in Core-Collapse Supernovae and Neutron Star Mergers}",
    eprint = "2503.05959",
    archivePrefix = "arXiv",
    primaryClass = "astro-ph.HE",
    reportNumber = "LA-UR-25-21809",
    doi = "10.1146/annurev-nucl-121423-100853",
    month = "3",
    year = "2025",
    journal = "Annu. Rev. Nucl. Part. Sci.",
    volume = "75"
}

@article{Fiorillo:2025npi,
    author = "Fiorillo, Damiano F. G. and Raffelt, Georg G.",
    title = "{Collective Flavor Conversions Are Interactions of Neutrinos with Quantized Flavor Waves}",
    eprint = "2502.06935",
    archivePrefix = "arXiv",
    primaryClass = "hep-ph",
    doi = "10.1103/PhysRevLett.134.211003",
    journal = "Phys. Rev. Lett.",
    volume = "134",
    number = "21",
    pages = "211003",
    year = "2025"
}

@article{Fiorillo:2024dik,
    author = "Fiorillo, Damiano F. G. and Goimil-Garc\'\i{}a, Manuel and Raffelt, Georg G.",
    title = "{Fast flavor pendulum: Instability condition}",
    eprint = "2412.09027",
    archivePrefix = "arXiv",
    primaryClass = "hep-ph",
    doi = "10.1103/PhysRevD.111.083028",
    journal = "Phys. Rev. D",
    volume = "111",
    number = "8",
    pages = "083028",
    year = "2025"
}

@article{Fiorillo:2025ank,
    author = "Fiorillo, Damiano F. G. and Raffelt, Georg G.",
    title = "{Theory of neutrino slow flavor evolution. Part II. Space-time evolution of linear instabilities}",
    eprint = "2501.16423",
    archivePrefix = "arXiv",
    primaryClass = "hep-ph",
    doi = "10.1007/JHEP06(2025)146",
    journal = "JHEP",
    volume = "06",
    pages = "146",
    year = "2025"
}

@article{Johns:2024bob,
    author = "Johns, Lucas",
    title = "{Implications of conservation laws and ergodicity for neutrino flavor instability}",
    eprint = "2402.08896",
    archivePrefix = "arXiv",
    primaryClass = "hep-ph",
    reportNumber = "LA-UR-24-21158",
    doi = "10.1103/qd8x-29zm",
    journal = "Phys. Rev. D",
    volume = "112",
    number = "6",
    pages = "063029",
    year = "2025"
}

@article{Yi:2019hrp,
    author = "Yi, Changhao and Ma, Lei and Martin, Joshua D. and Duan, Huaiyu",
    title = "{Dispersion relation of the fast neutrino oscillation wave}",
    eprint = "1901.01546",
    archivePrefix = "arXiv",
    primaryClass = "hep-ph",
    doi = "10.1103/PhysRevD.99.063005",
    journal = "Phys. Rev. D",
    volume = "99",
    number = "6",
    pages = "063005",
    year = "2019"
}

@article{Sawyer:2004ai,
    author = "Sawyer, R. F.",
    title = "{``Classical'' instabilities and ``quantum'' speed-up in the evolution of neutrino clouds}",
    eprint = "hep-ph/0408265",
    archivePrefix = "arXiv",
    month = "8",
    year = "2004"
}

@article{Sawyer:2008zs,
    author = "Sawyer, R. F.",
    title = "{The multi-angle instability in dense neutrino systems}",
    eprint = "0803.4319",
    archivePrefix = "arXiv",
    primaryClass = "astro-ph",
    doi = "10.1103/PhysRevD.79.105003",
    journal = "Phys. Rev. D",
    volume = "79",
    pages = "105003",
    year = "2009"
}

@article{Shalgar:2024gjt,
    author = "Shalgar, Shashank and Tamborra, Irene",
    title = "{Neutrino quantum kinetics in a core-collapse supernova}",
    eprint = "2406.09504",
    archivePrefix = "arXiv",
    primaryClass = "astro-ph.HE",
    doi = "10.1088/1475-7516/2024/09/021",
    journal = "JCAP",
    volume = "09",
    pages = "021",
    year = "2024"
}

@article{Cornelius:2024zsb,
    author = "Cornelius, Marie and Shalgar, Shashank and Tamborra, Irene",
    title = "{Neutrino quantum kinetics in two spatial dimensions}",
    eprint = "2407.04769",
    archivePrefix = "arXiv",
    primaryClass = "astro-ph.HE",
    doi = "10.1088/1475-7516/2024/11/060",
    journal = "JCAP",
    volume = "11",
    pages = "060",
    year = "2024"
}

@article{Fiorillo:2024bzm,
    author = "Fiorillo, Damiano F. G. and Raffelt, Georg G.",
    title = "{Theory of neutrino fast flavor evolution. Part~I. Linear response theory and stability conditions}",
    eprint = "2406.06708",
    archivePrefix = "arXiv",
    primaryClass = "hep-ph",
    doi = "10.1007/JHEP08(2024)225",
    journal = "JHEP",
    volume = "08",
    pages = "225",
    year = "2024"
}

@article{DedinNeto:2023ykt,
    author = "Dedin Neto, Pedro and Tamborra, Irene and Shalgar, Shashank",
    title = "{Energy Dependence of Flavor Instabilities Stemming from Crossings in the Neutrino Flavor Lepton Number Angular Distribution}",
    eprint = "2312.06556",
    archivePrefix = "arXiv",
    primaryClass = "astro-ph.HE",
    month = "12",
    year = "2023"
}

@article{Johns:2024dbe,
    author = "Johns, Lucas",
    title = "{Subgrid modeling of neutrino oscillations in astrophysics}",
    eprint = "2401.15247",
    archivePrefix = "arXiv",
    primaryClass = "astro-ph.HE",
    reportNumber = "LA-UR-24-20772",
    doi = "10.1103/3fr2-qttd",
    journal = "Phys. Rev. D",
    volume = "112",
    number = "4",
    pages = "043024",
    year = "2025"
}

@article{Duan:2006an,
    author = "Duan, Huaiyu and Fuller, George M. and Carlson, J and Qian, Yong-Zhong",
    title = "{Simulation of coherent nonlinear neutrino flavor transformation in the supernova environment: Correlated neutrino trajectories}",
    eprint = "astro-ph/0606616",
    archivePrefix = "arXiv",
    reportNumber = "LA-UR-06-4274",
    doi = "10.1103/PhysRevD.74.105014",
    journal = "Phys. Rev. D",
    volume = "74",
    pages = "105014",
    year = "2006"
}

@article{Capozzi:2017gqd,
    author = "Capozzi, Francesco and Dasgupta, Basudeb and Lisi, Eligio and Marrone, Antonio and Mirizzi, Alessandro",
    title = "{Fast flavor conversions of supernova neutrinos: Classifying instabilities via dispersion relations}",
    eprint = "1706.03360",
    archivePrefix = "arXiv",
    primaryClass = "hep-ph",
    reportNumber = "TIFR-TH-17-31",
    doi = "10.1103/PhysRevD.96.043016",
    journal = "Phys. Rev. D",
    volume = "96",
    number = "4",
    pages = "043016",
    year = "2017"
}

@article{Fiorillo:2024wej,
    author = {Fiorillo, Damiano F. G. and Raffelt, Georg G. and Sigl, G\"unter},
    title = "{Collective neutrino-antineutrino oscillations in dense neutrino environments?}",
    eprint = "2401.02478",
    archivePrefix = "arXiv",
    primaryClass = "hep-ph",
    doi = "10.1103/PhysRevD.109.043031",
    journal = "Phys. Rev. D",
    volume = "109",
    number = "4",
    pages = "043031",
    year = "2024"
}

@article{Dolgov:1980cq,
    author = "Dolgov, A. D.",
    title = "{Neutrinos in the early universe}",
    journal = "Sov. J. Nucl. Phys.",
    volume = "33",
    pages = "700--706",
    year = "1981",
    note = "[{\em Yad.\ Fiz.} {\bf 33} (1981) 1309]"
}

@article{Samuel:1993uw,
    author = "Samuel, Stuart",
    title = "{Neutrino oscillations in dense neutrino gases}",
    reportNumber = "IUHET-244",
    doi = "10.1103/PhysRevD.48.1462",
    journal = "Phys. Rev. D",
    volume = "48",
    pages = "1462--1477",
    year = "1993"
}

@article{Fiorillo:2024qbl,
    author = "Fiorillo, Damiano F. G. and Raffelt, Georg G.",
    title = "{Fast Flavor Conversions at the Edge of Instability in a Two-Beam Model}",
    eprint = "2403.12189",
    archivePrefix = "arXiv",
    primaryClass = "hep-ph",
    doi = "10.1103/PhysRevLett.133.221004",
    journal = "Phys. Rev. Lett.",
    volume = "133",
    number = "22",
    pages = "221004",
    year = "2024"
}

@article{Fiorillo:2024fnl,
    author = {Fiorillo, Damiano F. G. and Raffelt, Georg G. and Sigl, G\"unter},
    title = "{Inhomogeneous Kinetic Equation for Mixed Neutrinos: Tracing the Missing Energy}",
    eprint = "2401.05278",
    archivePrefix = "arXiv",
    primaryClass = "hep-ph",
    doi = "10.1103/PhysRevLett.133.021002",
    journal = "Phys. Rev. Lett.",
    volume = "133",
    number = "2",
    pages = "021002",
    year = "2024"
}

@article{Fiorillo:2024uki,
    author = "Fiorillo, Damiano F. G. and Raffelt, Georg G.",
    title = "{Theory of neutrino fast flavor evolution. Part II. Solutions at the edge of instability}",
    eprint = "2409.17232",
    archivePrefix = "arXiv",
    primaryClass = "hep-ph",
    doi = "10.1007/JHEP12(2024)205",
    journal = "JHEP",
    volume = "12",
    pages = "205",
    year = "2024"
}

@article{Nagakura:2022kic,
    author = "Nagakura, Hiroki and Zaizen, Masamichi",
    title = "{Time-Dependent and Quasisteady Features of Fast Neutrino-Flavor Conversion}",
    eprint = "2206.04097",
    archivePrefix = "arXiv",
    primaryClass = "astro-ph.HE",
    doi = "10.1103/PhysRevLett.129.261101",
    journal = "Phys. Rev. Lett.",
    volume = "129",
    number = "26",
    pages = "261101",
    year = "2022"
}

@article{Zaizen:2022cik,
    author = "Zaizen, Masamichi and Nagakura, Hiroki",
    title = "{Simple method for determining asymptotic states of fast neutrino-flavor conversion}",
    eprint = "2211.09343",
    archivePrefix = "arXiv",
    primaryClass = "astro-ph.HE",
    doi = "10.1103/PhysRevD.107.103022",
    journal = "Phys. Rev. D",
    volume = "107",
    number = "10",
    pages = "103022",
    year = "2023"
}

@article{Fiorillo:2023hlk,
    author = "Fiorillo, Damiano F. G. and Raffelt, Georg G.",
    title = "{Flavor solitons in dense neutrino gases}",
    eprint = "2303.12143",
    archivePrefix = "arXiv",
    primaryClass = "hep-ph",
    doi = "10.1103/PhysRevD.107.123024",
    journal = "Phys. Rev. D",
    volume = "107",
    number = "12",
    pages = "123024",
    year = "2023"
}

@article{Chakraborty:2016lct,
    author = "Chakraborty, Sovan and Hansen, Rasmus Sloth and Izaguirre, Ignacio and Raffelt, Georg",
    title = "{Self-induced neutrino flavor conversion without flavor mixing}",
    eprint = "1602.00698",
    archivePrefix = "arXiv",
    primaryClass = "hep-ph",
    doi = "10.1088/1475-7516/2016/03/042",
    journal = "JCAP",
    volume = "03",
    pages = "042",
    year = "2016"
}

@article{Tamborra:2020cul,
    author = "Tamborra, Irene and Shalgar, Shashank",
    title = "{New Developments in Flavor Evolution of a Dense Neutrino Gas}",
    eprint = "2011.01948",
    archivePrefix = "arXiv",
    primaryClass = "astro-ph.HE",
    doi = "10.1146/annurev-nucl-102920-050505",
    journal = "Ann. Rev. Nucl. Part. Sci.",
    volume = "71",
    pages = "165--188",
    year = "2021"
}

@article{Pantaleone:1992eq,
    author = "Pantaleone, James T.",
    title = "{Neutrino oscillations at high densities}",
    reportNumber = "DOE-ER-40561-056, INT-92-07-01",
    doi = "10.1016/0370-2693(92)91887-F",
    journal = "Phys. Lett. B",
    volume = "287",
    pages = "128--132",
    year = "1992"
}

@article{Padilla-Gay:2021haz,
    author = "Padilla-Gay, Ian and Tamborra, Irene and Raffelt, Georg G.",
    title = "{Neutrino Flavor Pendulum Reloaded: The Case of Fast Pairwise Conversion}",
    eprint = "2109.14627",
    archivePrefix = "arXiv",
    primaryClass = "astro-ph.HE",
    doi = "10.1103/PhysRevLett.128.121102",
    journal = "Phys. Rev. Lett.",
    volume = "128",
    number = "12",
    pages = "121102",
    year = "2022"
}

@article{Fiorillo:2023mze,
    author = "Fiorillo, Damiano F. G. and Raffelt, Georg G.",
    title = "{Slow and fast collective neutrino oscillations: Invariants and reciprocity}",
    eprint = "2301.09650",
    archivePrefix = "arXiv",
    primaryClass = "hep-ph",
    doi = "10.1103/PhysRevD.107.043024",
    journal = "Phys. Rev. D",
    volume = "107",
    number = "4",
    pages = "043024",
    year = "2023"
}

@article{Izaguirre:2016gsx,
    author = "Izaguirre, Ignacio and Raffelt, Georg and Tamborra, Irene",
    title = "{Fast Pairwise Conversion of Supernova Neutrinos: A Dispersion-Relation Approach}",
    eprint = "1610.01612",
    archivePrefix = "arXiv",
    primaryClass = "hep-ph",
    reportNumber = "INT-PUB-16-023, MPP-2016-266",
    doi = "10.1103/PhysRevLett.118.021101",
    journal = "Phys. Rev. Lett.",
    volume = "118",
    number = "2",
    pages = "021101",
    year = "2017"
}

@article{Sigl:1993ctk,
    author = "Sigl, G. and Raffelt, G.",
    title = "{General kinetic description of relativistic mixed neutrinos}",
    reportNumber = "MPI-PH-92-112",
    doi = "10.1016/0550-3213(93)90175-O",
    journal = "Nucl. Phys. B",
    volume = "406",
    pages = "423--451",
    year = "1993"
}

@ARTICLE{Rudsky,
       author = {{Rudzsky}, M.~A.},
        title = "{Kinetic equations for neutrino spin- and type-oscillations in a medium}",
      journal = {Astrophys. Space Sci},
         year = 1990,
        month = mar,
       volume = {165},
       number = {1},
        pages = {65-81},
          doi = {10.1007/BF00653658},
       adsurl = {https://ui.adsabs.harvard.edu/abs/1990Ap&SS.165...65R}
}

@article{Johns:2023jjt,
    author = "Johns, Lucas",
    title = "{Thermodynamics of oscillating neutrinos}",
    eprint = "2306.14982",
    archivePrefix = "arXiv",
    primaryClass = "hep-ph",
    reportNumber = "LA-UR-25-24518",
    doi = "10.1103/y8qs-w8lq",
    journal = "Phys. Rev. D",
    volume = "112",
    number = "6",
    pages = "063032",
    year = "2025"
}

@article{Johns:2021qby,
    author = "Johns, Lucas",
    title = "{Collisional Flavor Instabilities of Supernova Neutrinos}",
    eprint = "2104.11369",
    archivePrefix = "arXiv",
    primaryClass = "hep-ph",
    doi = "10.1103/PhysRevLett.130.191001",
    journal = "Phys. Rev. Lett.",
    volume = "130",
    number = "19",
    pages = "191001",
    year = "2023"
}

@article{Shalgar:2022rjj,
    author = "Shalgar, Shashank and Tamborra, Irene",
    title = "{Neutrino decoupling is altered by flavor conversion}",
    eprint = "2206.00676",
    archivePrefix = "arXiv",
    primaryClass = "astro-ph.HE",
    doi = "10.1103/PhysRevD.108.043006",
    journal = "Phys. Rev. D",
    volume = "108",
    number = "4",
    pages = "043006",
    year = "2023"
}

@article{Banerjee:2011fj,
    author = "Banerjee, Arka and Dighe, Amol and Raffelt, Georg",
    title = "{Linearized flavor-stability analysis of dense neutrino streams}",
    eprint = "1107.2308",
    archivePrefix = "arXiv",
    primaryClass = "hep-ph",
    reportNumber = "MPP-2011-81, TIFR-TH-11-30",
    doi = "10.1103/PhysRevD.84.053013",
    journal = "Phys. Rev. D",
    volume = "84",
    pages = "053013",
    year = "2011"
}

@article{Morinaga:2021vmc,
    author = "Morinaga, Taiki",
    title = "{Fast neutrino flavor instability and neutrino flavor lepton number crossings}",
    eprint = "2103.15267",
    archivePrefix = "arXiv",
    primaryClass = "hep-ph",
    doi = "10.1103/PhysRevD.105.L101301",
    journal = "Phys. Rev. D",
    volume = "105",
    number = "10",
    pages = "L101301",
    year = "2022"
}

@article{Raffelt:2007cb,
    author = "Raffelt, Georg G. and Smirnov, Alexei {\relax{Yu}}",
    title = "{Self-induced spectral splits in supernova neutrino fluxes}",
    eprint = "0705.1830",
    archivePrefix = "arXiv",
    primaryClass = "hep-ph",
    reportNumber = "MPP-2007-53",
    doi = "10.1103/PhysRevD.76.081301",
    journal = "Phys. Rev. D",
    volume = "76",
    pages = "081301",
    year = "2007",
    note = "Erratum:
    \href{https://doi.org/10.1103/PhysRevD.77.029903}{{\em Phys. Rev. D} {\bf 77} (2008) 029903}"
}

@article{Samuel:1995ri,
    author = "Samuel, Stuart",
    title = "{Bimodal coherence in dense selfinteracting neutrino gases}",
    eprint = "hep-ph/9604341",
    archivePrefix = "arXiv",
    reportNumber = "MPI-PHT-95-57, CCNY-HEP-95-5",
    doi = "10.1103/PhysRevD.53.5382",
    journal = "Phys. Rev. D",
    volume = "53",
    pages = "5382--5393",
    year = "1996"
}

@article{Duan:2010bg,
    author = "Duan, Huaiyu and Fuller, George M. and Qian, Yong-Zhong",
    title = "{Collective Neutrino Oscillations}",
    eprint = "1001.2799",
    archivePrefix = "arXiv",
    primaryClass = "hep-ph",
    reportNumber = "LA-UR-09-08309, INT-PUB-10-001",
    doi = "10.1146/annurev.nucl.012809.104524",
    journal = "Ann. Rev. Nucl. Part. Sci.",
    volume = "60",
    pages = "569--594",
    year = "2010"
}

@article{Wolfenstein:1979ni,
    author = "Wolfenstein, L.",
    title = "{Neutrino Oscillations and Stellar Collapse}",
    reportNumber = "COO-3066-133",
    doi = "10.1103/PhysRevD.20.2634",
    journal = "Phys. Rev. D",
    volume = "20",
    pages = "2634--2635",
    year = "1979"
}

@article{Johns:2019izj,
    author = "Johns, Lucas and Nagakura, Hiroki and Fuller, George M. and Burrows, Adam",
    title = "{Neutrino oscillations in supernovae: angular moments and fast instabilities}",
    eprint = "1910.05682",
    archivePrefix = "arXiv",
    NoprimaryClass = "hep-ph",
    doi = "10.1103/PhysRevD.101.043009",
    journal = "Phys. Rev. D",
    volume = "101",
    number = "4",
    pages = "043009",
    year = "2020"
}

\end{document}